%% file: main.tex
\documentclass[conference]{IEEEtran}
\usepackage{cite}
\usepackage{amsmath,amssymb,amsfonts}
\usepackage{amsthm}
\usepackage{comment}
\usepackage{booktabs}
\usepackage{textcomp}
\usepackage{xcolor}
\def\BibTeX{{\rm B\kern-.05em{\sc i\kern-.025em b}\kern-.08em
    T\kern-.1667em\lower.7ex\hbox{E}\kern-.125emX}}

\usepackage{algorithm}
\usepackage{algcompatible}
\usepackage[noend]{algpseudocode} 
\usepackage{tikz}
\usetikzlibrary{calc}
\usetikzlibrary{shapes.geometric, arrows.meta, positioning}

\DeclareRobustCommand{\halfcheckmark}{%
  \tikz[x=1em,y=1em,baseline=-0.1em]{%
    \fill (0,.35) -- (.25,0) -- (1,.7) -- (.25,.15) -- cycle;
    \fill (.60,.70) -- (.65,.70) -- (.82,.20) -- (.77,.20) -- cycle;
  }%
}

\usepackage{array}

\usepackage{caption}
\usepackage{graphicx}
\usepackage{subcaption}

\usepackage{placeins} 

\usepackage{enumitem}
\setlist[itemize]{leftmargin=*,itemsep=0pt, topsep=0pt, parsep=0pt, partopsep=0pt}
\setlist[enumerate]{leftmargin=*,itemsep=0pt, topsep=0pt, parsep=0pt, partopsep=0pt}

\usepackage{etoolbox}
\newtoggle{TechnicalReport}
\togglefalse{TechnicalReport}
\newtoggle{LinkTechnicalReport}
\togglefalse{LinkTechnicalReport}

\newtheorem{theorem}{Theorem}[section]
\newtheorem{lemma}[theorem]{Lemma}
\theoremstyle{definition}
\newtheorem{definition}[theorem]{Definition}
\newtheorem{example}[theorem]{Example}
\theoremstyle{plain}

\usepackage[hidelinks]{hyperref}
\newcommand{\technicalreport}{\iftoggle{LinkTechnicalReport}{\href{https://github.com/JunyuLiao/High-dimensional-Regret-Minimization/blob/main/docs/TechnicalReport.pdf}{technical report}}{technical report}}

\begin{document}

\title{High-dimensional Regret Minimization}

\author{\IEEEauthorblockN{\makebox[0.48\textwidth]{Junyu Liao}}
\IEEEauthorblockA{
\textit{The Hong Kong University of Science and Technology}\\
jliaoak@connect.ust.hk}
\and
\IEEEauthorblockN{\makebox[0.48\textwidth]{Ashwin Lall}}
\IEEEauthorblockA{
\textit{Denison University}\\
lalla@denison.edu}
\and[\hfill\mbox{}\par\vspace{1em}\mbox{}\hfill]
\IEEEauthorblockN{\makebox[0.48\textwidth]{Mitsunori Ogihara}}
\IEEEauthorblockA{
\textit{University of Miami}\\
m.ogihara@miami.edu}
\and
\IEEEauthorblockN{\makebox[0.48\textwidth]{Raymond Chi-Wing Wong}}
\IEEEauthorblockA{
\textit{The Hong Kong University of Science and Technology}\\
raywong@cse.ust.hk}
}

\maketitle

\begin{abstract}
\input{abstract}

\end{abstract}

\begin{IEEEkeywords}
regret minimization, user interaction, multi-criteria decision making
\end{IEEEkeywords}

\input{introduction}

\input{related}
\input{definitions}

\input{algorithm}
\input{attributesubset}
\input{experiment}

\input{conclusions}

\bibliographystyle{IEEEtran}
\bibliography{references}

\input{appendix}

\end{document}

%% file: abstract.tex
Multi-criteria decision making over large databases is fundamental to many real-world applications. Interactive regret minimization has emerged recently as a promising approach that combines the controllable output size of top-$k$ queries with the flexibility of skyline queries, without requiring users to explicitly specify a utility function. Instead, the system iteratively presents a small set of options, learns the user’s implicit preferences from their selections, and progressively refines its recommendations.

However, modern applications such as housing search and financial product selection often involve hundreds of attributes. Existing interactive methods either fail to scale to such high-dimensional data or require excessive user interaction, frequently exceeding $1000$ rounds. We address this limitation with \textsc{FHDR} (\underline{F}ast \underline{H}igh-\underline{D}imensional \underline{R}eduction), a framework that exploits sparse linear preferences to discover key attributes through lightweight interactions. We also propose \textsc{AttributeSubset}, an efficient single-round algorithm that returns a compact recommendation set without further feedback. In our experiments on real datasets, \textsc{FHDR} completes in under $0.03$s and requires fewer than $30$ interaction rounds under the default setting.

Extensive experiments on both synthetic and real-world datasets show that \textsc{FHDR} outperforms the best-known baselines by at least one order of magnitude in execution time and by up to several orders of magnitude in interaction cost. These results establish \textsc{FHDR} as a new state of the art for scalable interactive regret minimization.

%% file: introduction.tex
\section{Introduction}
\label{sec:intro}

Multi-criteria decision making over large databases is important in many real-world applications.
A common approach to facilitating user decision-making in databases is to model user preference as a \textit{utility function} \cite{CB00, NL12, NS10}, which assigns each tuple a numerical score. Knowledge of this function—either explicit or implicit—enables systems to reduce user effort in identifying preferred tuples. Two major paradigms built on this assumption are (1) the \emph{top-$k$ query} \cite{GY05, PL07, SI07, LYH09, LC09} and (2) the \emph{skyline query} \cite{CJ06, CJ062, LY07, MC09, PT05}.
(1) In a top-$k$ query, the user explicitly provides a utility function and an integer $k$, and the system returns the $k$ highest-scoring tuples. While conceptually straightforward, this approach is impractical when the user cannot specify precise weights in the utility function—especially in high-dimensional data, where hundreds of attributes make this specification infeasible.
(2) The skyline query, by contrast, assumes only that the utility function is component-wise monotonic: if one tuple is no worse than another in every attribute, its utility is at least as high. The system thus returns all skyline points, i.e., tuples not dominated by any other. Although this avoids explicit utility specification, the output can become unmanageably large. As dimensionality increases, domination becomes less common and the skyline typically contains more tuples, potentially overwhelming the user.

Recently, the $k$-regret query~\cite{AK17, asudeh2017efficient, NS10, PW14, XW18} has emerged as a promising solution to the limitations of top-$k$ and skyline queries. The method selects $k$ representative tuples such that, for any possible user preference, the best tuple in the returned set closely approximates the global optimum in the database. The regret ratio measures the relative utility loss incurred by choosing from the returned set rather than the entire database (Section~\ref{sec:def}). The objective is to minimize the maximum regret ratio over the admissible utility functions.

These frameworks operate in a single round, presenting their results at once. This naturally raises the question: can multiple rounds of interaction further reduce regret or output size? To address this, interactive regret minimization frameworks~\cite{NL12, XCL19} have been proposed, where the system iteratively presents a few candidate tuples per round, learns the user’s implicit utility function from their choices, and progressively refines subsequent recommendations.

Most previous studies have focused on databases with a small number of attributes, typically 2-10, and often require users to specify their attributes of interest beforehand. In contrast, this paper considers a \textit{high-dimensional setting} with more than a hundred attributes and no prior attribute selection. These scenarios are increasingly common nowadays, e.g., in housing markets (price, size, location, etc.) and financial products (interest rate, liquidity, historical performance, etc.). In these domains, the high dimensionality poses significant challenges: computational costs grow exponentially, theoretical guarantees become weaker, and traditional algorithms often fail to provide meaningful representative subsets. There is an urgent need for efficient and effective algorithms that address these challenges.

We want algorithms that satisfy the following requirements:

\begin{itemize}[beginpenalty=10000]
    \item \textit{Quality.} Return a small, manageable set with low regret, ideally identifying the user’s favorite tuple in the database.
    
    \item \textit{Efficiency.} Execute within a reasonable time on high-dimensional datasets that may contain millions of tuples.
    
    \item \textit{Usability.} Require few interactions, using simple, intuitive questions that remain manageable as dimensionality grows.
\end{itemize}

Our work is motivated by the observation that users typically base decisions on a small subset of salient attributes rather than the full feature set. For example, a house may be described by hundreds of attributes, including floor area, nearby amenities, and energy efficiency, yet buyers often focus primarily on a few factors such as price, size, and location. This behavior is well documented in cognitive psychology~\cite{GG96, GA12, MH99, GG99} and conjoint analysis~\cite{WX22, SC05, TK03, DMW12}, and is further supported by our survey.

In psychology, the “Take-The-Best” heuristic~\cite{GG96, GA12} shows that people make accurate decisions using only a few salient attributes. It ranks attributes by importance and bases each choice on the first attributes that distinguish the alternatives, ignoring the rest. Despite its simplicity, it can outperform more complex models, including multiple regression and weighted integration~\cite{MH99, GG99}. This finding supports our high-dimensional $k$-regret setting and motivates the assumption that the true utility vector is sparse, with most weight concentrated on a small subset of attributes.

Conjoint analysis similarly estimates preferences over multiple attributes from evaluations or choices among alternatives~\cite{WX22, SC05, TK03, DMW12}. Estimated attribute contributions yield relative importance scores, providing evidence that importance can be concentrated on a small subset of attributes. In Schaupp and Bélanger's study, the five highest-ranked attributes account for $84.2\%$ of the reported relative importance~\cite{SC05}. This concentration of importance motivates our sparse-utility model.

\begin{figure}[h]
    \centering
    \includegraphics[width=0.7\linewidth]{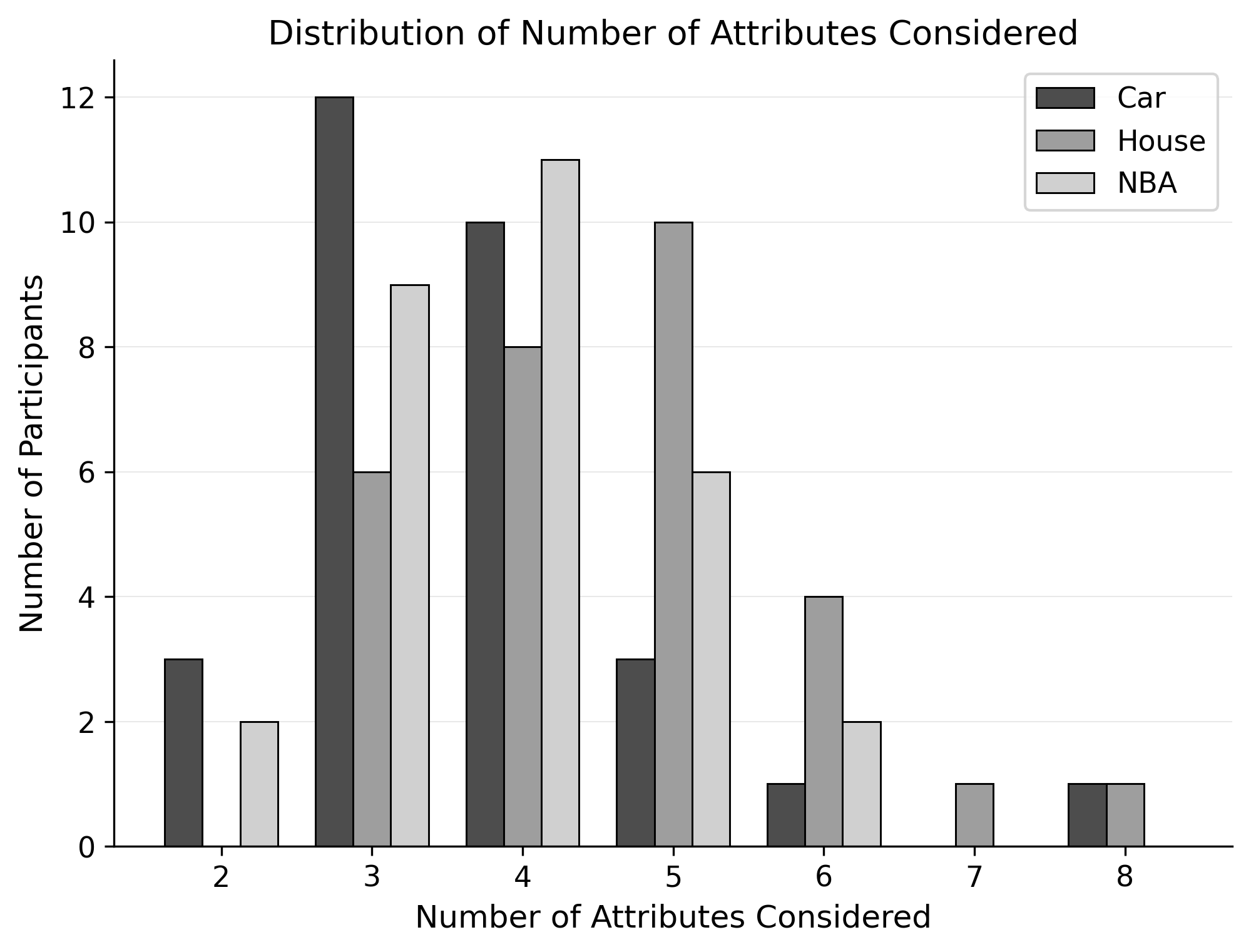}
    \caption{Survey on Number of Attributes Considered}
    \label{fig:num_att}
\end{figure}

To examine how many attributes users consider in decision making, we surveyed $30$ participants using the Car, House, and NBA datasets described in Section~\ref{sec:eval}. Participants were shown a set of options and asked to identify the attributes they considered when comparing them. They could select any number of attributes, with no minimum or maximum imposed. The datasets contain $57$, $33$, and $104$ attributes, respectively. Figure~\ref{fig:num_att} summarizes the number of attributes selected for each dataset and shows that most participants relied on only a small subset. For the Car dataset, $12$ participants ($40\%$) considered exactly $3$ of the $57$ attributes, with \textit{Real-World MPG}, \textit{Model Year}, and \textit{Engine Displacement} being the three most frequently selected. Across all datasets, more than $80\%$ of participants considered at most $5$ attributes. These results provide empirical support for our assumption that users' effective utility functions are sparse.

Motivated by this, we introduce FHDR (\underline{F}ast \underline{H}igh-\underline{D}imensional \underline{R}eduction), a generic framework for $k$-regret minimization in high-dimensional settings. Our major contributions are as follows:
\begin{itemize}
    \item We present FHDR, a framework for high-dimensional regret minimization under sparse linear preferences. It handles datasets of arbitrary original dimensionality and identifies the user's favorite tuple with sufficient feedback, requiring substantially fewer questions in the evaluated comparisons.
    \item We implicitly discover the user's key attributes through tuple comparisons, avoiding upfront attribute selection requirement in previous papers.
    \item We support flexible interaction, including zero feedback. The user can skip or stop answering questions at any time, ensuring a user-friendly interaction process.
    \item We propose a single-round algorithm \textsc{AttributeSubset} for any dimensionality, which is more than $10$ times faster than existing single-round algorithms under high-dimensional settings.
    \item We conduct extensive experiments on synthetic and real datasets to evaluate regret ratio, computation time, and interaction cost. A user study on Car further evaluates FHDR's end-to-end recommendation quality and usability.
\end{itemize}

The rest of the paper is organized as follows: Section~\ref{sec:related} discusses related work, and Section~\ref{sec:def} gives the problem definition. The complete algorithm for high-dimensional regret minimization is presented in Section~\ref{sec:alg}, while the solution for limited user feedback is given in Section~\ref{sec:attsub}. Section~\ref{sec:eval} shows the experiment results and Section~\ref{sec:concl} concludes the paper.

%% file: related.tex
\section{Related Work}
\label{sec:related}

The $k$-regret query~\cite{NS10}, also known as the regret-minimizing set query, addresses the limitations of top-$k$ and skyline queries. It returns $k$ representative tuples such that, for any unknown nonnegative linear utility function over $d$ attributes, the best returned tuple closely approximates the global optimum. Its objective is to minimize the maximum regret ratio defined in Section~\ref{sec:def}, thereby avoiding explicit utility specification while controlling the output size. However, finding an optimal $k$-regret set is NP-hard~\cite{AK17, CL17, CT14}, particularly in high dimensions. Early methods, including \textsc{Cube}~\cite{NS10}, \textsc{Greedy}~\cite{NS10}, and \textsc{GeoGreedy}~\cite{PW14}, either lack theoretical guarantees or perform poorly in practice. Later approaches such as \textsc{HD-Greedy}~\cite{asudeh2017efficient} and $\varepsilon$-\textsc{Kernel}~\cite{CL17} improve efficiency but offer weak bounds: \textsc{HD-Greedy} provides only a loose upper bound, while the bound of $\varepsilon$-\textsc{Kernel} may exceed $1$ even for $d<10$, making it uninformative in practice.

Recent advances have extended regret minimization to high-dimensional settings. Xie et al.~\cite{XW18} proposed the \textsc{Sphere} algorithm, which achieves an asymptotically optimal upper bound on the maximum regret ratio without restricting data dimensionality. Although earlier methods such as Cube ~\cite{NS10} also provide theoretical guarantees that apply across dimensions, Sphere is distinguished by offering an asymptotically optimal bound while also demonstrating strong empirical performance, making it a state-of-the-art single-round solution. However, its performance degrades as dimensionality increases due to two limitations: (1) it requires the output size $k$ to be at least the dimensionality $d$, which is impractical in high-dimensional applications, and (2) its runtime grows sub-exponentially with $d$, rendering it inefficient for large datasets. These issues motivate the development of complementary approaches for scalable high-dimensional decision making.

Another line of work studies interactive regret minimization, which replaces one-shot query processing with multi-round interaction to progressively infer the user’s implicit utility function. The first such framework, \textsc{UtilityApprox}~\cite{NL12}, repeatedly presents a small set of tuples and asks the user to select the preferred one. Each response refines the estimated utility function, allowing subsequent queries to focus on more promising tuples and reduce regret for a fixed output size~\cite{NL12}. However, \textsc{UtilityApprox} relies on synthetic tuples, which may frustrate users when a preferred option is not real, and requires an impractically large number of questions in high-dimensional settings.

Xie et al.~\cite{XCL19} addressed this limitation with a strongly truthful framework that displays only real tuples, improving user trust and simplifying preference elicitation. Their algorithms, including \textsc{UH-Random} and \textsc{UH-Simplex}, converge to near-optimal results within a bounded number of rounds and achieve an asymptotically optimal question complexity in two dimensions. However, their scalability is limited in higher dimensions by numerical instability in convex-hull computations. To overcome this issue, these methods require users to explicitly specify a small subset of relevant attributes, which imposes substantial burden in high-dimensional settings, as will be justified in later paragraphs.

Several other interactive methods have also been proposed. \textsc{Active-Ranking}~\cite{jamieson2011active} adaptively selects pairwise comparisons to recover a complete ranking under a Euclidean embedding assumption. However, it targets full-ranking recovery rather than regret minimization and assumes that the data are in general position, which may not hold in practical databases. \textsc{Preference-Learning}~\cite{qian2015learning} similarly reconstructs the entire utility function through adaptive comparisons, requiring queries that may be irrelevant to identifying the user’s preferred tuples. More recently, Wang et al.~\cite{WWX21} proposed \textsc{HD-PI} and \textsc{RH} for identifying top-$k$ tuples. Although these methods provide dimension-independent guarantees under ideal conditions, their scalability deteriorates as dimensionality increases because convex-hull and skyline frontiers grow rapidly, increasing both computational and interaction costs. Chen and Wong~\cite{CW23} further study best-tuple retrieval under random interaction errors through \textsc{Verify-Point} and \textsc{Verify-Space}. These methods provide probabilistic retrieval guarantees over a predefined attribute space, whereas FHDR discovers key attributes implicitly from high-dimensional data.

A further limitation of existing works is upfront attribute selection. For example, a user may be asked to select $5$ attributes from $100$ directly, before seeing actual candidate tuples. Do et al.~\cite{LA25} simplify interaction by comparing two attributes at a time, but still require users to select their attributes of interest. Prior work often treats this step as trivial preprocessing, while our survey and user study indicate that upfront attribute selection can be burdensome.

This selection can impose a substantial cognitive burden: users must inspect and compare many attributes while working memory is limited~\cite{Cowan_2001}. Moreover, users’ preferences are often constructed during actual decision-making rather than specified in advance~\cite{Slovic1995}, so selection without concrete candidate tuples risks prematurely excluding relevant attributes. On the other hand, partial-profile designs, which show only a subset of attributes at a time, have been reported to reduce decision complexity and choice randomness while improving response efficiency~\cite{meyerhoff_performance_2023}. Comparisons between concrete tuples also offer a more intuitive way to elicit preferences~\cite{NL12, XCL19, WWX21, CW23}. For example, a user may be unable to state abstractly whether ``location'' is globally more important than ``floor area'', yet can readily choose between two houses once the trade-off is instantiated. In this way, comparison-based interaction reveals preferences implicitly with less cognitive effort. These observations motivate FHDR’s implicit discovery of key attributes through tuple comparisons.

Compared with existing works, FHDR natively supports high-dimensional regret minimization and maintains strong performance as dimensionality increases, addressing key scalability barriers. Its user-friendly design requires no upfront attribute selection and allows users to skip or stop answering questions at any time. Table~\ref{tab:capabilities} summarizes the advantages of our method, alongside comparison with existing methods. By combining high-dimensional scalability, computational efficiency, and flexible interaction, FHDR advances the state of the art and enables practical and accurate decision support for complex, real-world multi-criteria applications.

\begin{table*}[!t]
  \centering
  \small
  \setlength{\tabcolsep}{3pt}
  \renewcommand{\arraystretch}{1.15}
  \begin{tabular}{@{}p{0.23\textwidth}*{5}{>{\centering\arraybackslash}p{0.14\textwidth}}@{}}
    \toprule
    \textbf{Method} & \textbf{Practical high-dimensional support} & \textbf{Exact-favorite guarantee} & \textbf{Implicit key-attribute discovery} & \textbf{User-controlled output size} & \textbf{Output at any interaction budget} \\
    \midrule
    \textsc{Sphere}~\cite{XW18} & \halfcheckmark & $\times$ & $\times$ & $\times$ & $\checkmark$ \\
    \textsc{UtilityApprox}~\cite{NL12} & \halfcheckmark & $\checkmark$ & $\times$ & $\times$ & $\times$ \\
    \textsc{UH-Random}~\cite{XCL19} & $\times$ & $\checkmark$ & $\times$ & $\times$ & $\times$ \\
    \textsc{HD-PI} / \textsc{RH}~\cite{WWX21} & $\times^{\dagger}$ & $\checkmark$ & $\times$ & $\checkmark$ & $\times$ \\
    \textsc{Verify-Point}~\cite{CW23} & $\times^{\dagger}$ & $\checkmark$ & $\times$ & $\checkmark$ & $\times$ \\
    \textsc{Breakpoint}~\cite{LA25} & \halfcheckmark & $\times$ & $\times$ & $\times$ & $\checkmark$ \\
    \midrule
    \textbf{FHDR} & $\checkmark$ & $\checkmark$ & $\checkmark$ & $\checkmark$ & $\checkmark$ \\
    \bottomrule
  \end{tabular}
  \caption{Capabilities of representative methods. Practical high-dimensional support reflects practical computation and interaction costs in the evaluated high-dimensional settings; \halfcheckmark\ denotes partial support with substantial practical limitations. $\dagger$ Execution entries concern the tested representatives \textsc{HD-PI} and \textsc{Verify-Point}, respectively. Detailed qualifications and reproduction results appear in \iftoggle{TechnicalReport}{Section~\ref{appen:capabilities} of the \technicalreport}{Appendix~\ref{appen:capabilities}}.}
  \label{tab:capabilities}
\end{table*}

%% file: definitions.tex
\section{Problem Definition}
\label{sec:def}

The input is a preprocessed dataset $X\subseteq[0,1]^d$ of $n$ points. Here, $d$ denotes the number of attributes. We assume that a user is only interested in $d_\text{int}$ among all $d$ attributes, where $d_\text{int}$ is a small number compared to $d$ (e.g., $d_\text{int}=3$ and $d=100$).

\subsection{Terminology}

We refer to a database object as ``\textit{tuple}" and ``\textit{point}" interchangeably throughout the paper. We denote the $i$-th dimensional value of a point $p$ by $p[i]$ where $i \in [1,d]$. For an attribute set $\mathcal{D}$, let $p_{\mathcal{D}}$ denote the projection of $p$ onto $\mathcal{D}$, and let $X_{\mathcal{D}}=\{p_{\mathcal{D}}:p\in X\}$. We remove constant attributes and apply \textit{min-max} normalization to each remaining attribute. Let $a_i$ and $b_i$ denote the minimum and maximum raw values of attribute $i$. A raw value $x[i]$ is mapped to $p[i]=(x[i]-a_i)/(b_i-a_i)$ when larger values are preferred, and to $p[i]=(b_i-x[i])/(b_i-a_i)$ when smaller values are preferred, as with \textit{price}. Thus, every retained attribute ranges from $0$ to $1$, attains both endpoints, and has larger normalized values preferred. Utilities and regret ratios are defined on this normalized representation. As an example, Table~\ref{tab:house} illustrates a database $X=\{p_1, p_2, p_3, p_4, p_5\}$ of $5$ points ($n=5$) where each point is described by $5$ attributes ($d=5$).

\begin{table}[!t]
\centering
\begin{tabular}{|c|c|c|c|c|c|c|}
\hline
\textbf{House} & $D_1$ & $D_2$ & $D_3$ & $D_4$ & $D_5$ & \textbf{Utility} $f(p)$ \\
\hline
$p_1$ & 0.84 & 0.61 & 0.93 & 0.70 & 0.31 & 0.782 \\
\hline
$p_2$ & 0.59 & 0.95 & 0.77 & 0.86 & 0.79 & 0.761 \\
\hline
$p_3$ & 0.69 & 0.84 & 1.00 & 0.99 & 0.55 & 0.820 \\
\hline
$p_4$ & 1.00 & 0.64 & 0.68 & 0.45 & 1.00 & 0.794 \\
\hline
$p_5$ & 0.74 & 1.00 & 0.44 & 1.00 & 0.73 & 0.756 \\
\hline
\end{tabular}
\caption{House dataset with five attributes and utility values for
$u=(0.40,0.35,0.25,0,0)$ ($d_\text{int}=3$).
The attributes $D_1\!\sim\!D_5$ denote \textit{price}, \textit{size}, \textit{commute time}, \textit{age} and \textit{condition score}, respectively.}
\label{tab:house}
\end{table}

Following~\cite{CB00, LC09, NL12, NS10, PW14}, we model user preferences by an unknown nonnegative linear utility function $f(p)=u\cdot p$, where $u\in\mathbb{R}_{\geq0}^{d}$ and $u[i]$ represents the importance of attribute $i$. We use $f$ and its utility vector $u$ interchangeably.

Building on prior studies that individuals consider only a limited number of attributes in high-dimensional decision making tasks~\cite{GG96, GA12, WX22, SC05, TK03, DMW12}, along with our own user survey, we assume that $u$ has exactly $d_\text{int}$ positive entries, where $d_\text{int}\ll d$. From findings from previous studies and a user survey, we adopt an upper bound $d_\text{max}$ of $5$ (i.e., $d_\text{int}\leq d_\text{max}=5$). An attribute (dimension) is said to be \textit{key} if the coefficient of that attribute in the utility vector $u[i]>0$; and \textit{non-key} otherwise.

The candidate attribute set $\mathcal{D}_\text{cand}$ contains all attributes not yet identified as non-key. Here, `candidate' refers to attributes rather than tuples, to avoid confusion.

We define the \textit{regret ratio}\cite{NS10} as follows:

\begin{definition}\cite{NS10}
Given a nonempty set $S \subseteq X$ and a utility vector $u$, the \emph{regret ratio} of $S$ over $X$ w.r.t.\ $u$, denoted by $\mathrm{rr}_X(S, u)$, is defined as
\[
\frac{\max_{p \in X} u \cdot p - \max_{p \in S} u \cdot p}{\max_{p \in X} u \cdot p}
= 1 - \frac{\max_{p \in S} u \cdot p}{\max_{p \in X} u \cdot p}.
\]
\end{definition}

The regret ratio is invariant under positive scaling of $u$. We therefore normalize $u$ so that $\sum_{i=1}^{d}u[i]=1$ ($\lVert u\rVert_1=1$).

For any nonempty subset $S\subseteq X$, the regret ratio lies between $0\%$ and $100\%$. Since each retained attribute attains $1$ and $\lVert u\rVert_1=1$, we have $\max_{p\in X}u\cdot p\geq\max_i u[i]>0$, so the regret ratio is well defined. A lower regret ratio means that the best returned tuple has utility closer to the dataset optimum. A tuple $p^*\in\arg\max_{p\in X}u\cdot p$ is a \textit{favorite tuple}, also called an optimal or maximum-utility tuple.

Consider the example given in Table~\ref{tab:house}. We have the utility vector $u = (0.40, 0.35, 0.25, 0, 0)$. Take the first point $p_1$ for illustration. Its corresponding utility w.r.t.\ $u$ can be calculated as: $f(p_1) = u \cdot p_1 = 0.782$. The utilities of other points are also presented in Table~\ref{tab:house}. The favorite tuple in $X$ is $p_3$, with utility $f(p_3) = 0.820$. If $S = \{p_1, p_2\}$, then the regret ratio of $S$ over $X$ w.r.t.\ $u$ is:
\[
\begin{aligned}
\mathrm{rr}_X(S, u) &= 1 - \frac{\max_{p \in S} u \cdot p}{\max_{p \in X} u \cdot p} = 1-\frac{0.782}{0.820} \approx 4.63\%.
\end{aligned}
\]

Notation is summarized in \iftoggle{TechnicalReport}{Section~\ref{appen:summary} of the \technicalreport}{Appendix~\ref{appen:summary}}.

\subsection{Interaction}

Following interactive regret-minimization frameworks~\cite{NL12, XCL19, WWX21, CW23}, FHDR learns preferences through tuple comparisons. It discovers key attributes and refines their relative utility weights without requiring explicit utility specification.

\paragraph{Format}
Each question presents $s$ tuples from $X$, denoted $\{p_1,p_2,\dots,p_s\}$, described only by attributes in $\mathcal{D}\subseteq\{D_1,D_2,\dots,D_d\}$, where $1\leq\lvert\mathcal{D}\rvert\leq m$. Here, $m$ is the maximum number of attributes displayed per question. The user either selects the most preferred tuple based solely on the displayed attributes or indicates that none of these attributes matters. We call the latter an \textit{irrelevant-attributes response}. This response is distinct from stopping the interaction.

\paragraph{Strong truthfulness.}
Following \cite{XCL19}, an interactive algorithm is \emph{strongly truthful} if it presents only existing objects to the user throughout the interaction. Our algorithm satisfies strong truthfulness because every tuple shown to the user is drawn from the original dataset.

\paragraph{Partial Utility Assumption}\label{par}
We assume a fixed utility vector $u$ and user responses consistent with the partial-utility and attribute-relevance rules below. Users are instructed to treat omitted attributes as identical for each comparison~\cite{chan2024vaccine}, so their common contribution cancels. Following Do et al.~\cite{LA25}, comparisons use the \textit{partial utility} $f_\mathcal{D}(p)=\sum_{i:D_i\in\mathcal{D}}u[i]p[i]$, retaining the original coefficients of the displayed attributes. Users select a tuple maximizing partial utility when at least one displayed attribute is key; otherwise, they give an irrelevant-attributes response.

For the utility vector in Table~\ref{tab:house}, displaying $\mathcal{D}=\{D_1,D_2,D_4\}$ gives $f_\mathcal{D}(p)=0.40p[1]+0.35p[2]$, since $u[4]=0$. The five partial utilities are $0.550$, $0.569$, $0.570$, $0.624$, and $0.646$, respectively, so the user selects $p_5$. If only $D_4$ and $D_5$ are displayed, both weights are zero and the user gives an irrelevant-attributes response.

\paragraph{Information Gained.}
Each response conveys information about the relevance of attributes in $\mathcal{D}$:
\begin{itemize}
    \item A tuple selection indicates that at least one displayed attribute is key.
    \item An irrelevant-attributes response indicates that $u[i]=0$ for every displayed attribute $D_i\in\mathcal{D}$.
\end{itemize}
These responses refine $\mathcal{D}_\text{cand}$. A tuple selection also supplies comparisons on displayed coordinates that can constrain relative utility weights.

\paragraph{Early Termination.}
Users may stop at any time. Before interactive Phase 3, including at zero feedback, \textsc{AttributeSubset} returns a recommendation set using the current candidate attributes. During interactive Phase 3, early stopping returns one recommendation based on the current feasible utility region; exact favorite-tuple retrieval requires sufficient feedback.

\subsection{Objectives}

We consider two objectives based on feedback.

\paragraph{Problem 1: }\label{p1}
With sufficient feedback under the stated response model, our primary objective is to return a favorite tuple $p^*\in\arg\max_{p\in X}u\cdot p$.

\paragraph{Problem 2: }\label{p2}
When interaction ends before interactive Phase 3, we seek a subset $S\subseteq X$ of user-specified size $\lvert S\rvert=K$ with a small regret ratio $\mathrm{rr}_X(S,u)$, without further feedback~\cite{NS10}. This objective includes the zero-feedback case.

%% file: algorithm.tex
\section{Algorithm}
\label{sec:alg}

We present \textsc{FHDR} for high-dimensional regret minimization under sparse linear preferences. The framework supports favorite-tuple retrieval with sufficient feedback (Problem~1~[\ref{p1}]) and a compact recommendation set under limited feedback (Problem~2~[\ref{p2}]). Section~\ref{subsec:framework} introduces the overall framework, Section~\ref{subsec:dimR} describes attribute discovery, and Section~\ref{subsec:re_min} presents the final output selection. Section~\ref{subsec:sum} summarizes the complete procedure. Detailed proofs are provided in \iftoggle{TechnicalReport}{Section~\ref{appen:proof} of the \technicalreport}{Appendix~\ref{appen:proof}}.

\subsection{Overall Framework}
\label{subsec:framework}

Phases 1 and 2 exploit sparse preferences to discover relevant attributes without requiring upfront attribute selection. Phase 3 refines preferences over the retained attributes. Throughout discovery, we maintain a candidate set $\mathcal{D}_\text{cand}$, initialized with all attributes, and remove attributes identified as non-key.

\paragraph{Phase 1: Coarse Elimination.}
Let $m\geq1$ be an integer specifying the maximum number of attributes displayed per question. Starting with all $d$ attributes, we partition them into blocks of at most $m$ attributes. Each block is tested through a tuple comparison under the protocol in Section~\ref{sec:def}. An irrelevant-attributes response eliminates the block; a tuple selection retains it for Phase 2.

\paragraph{Phase 2: Fine Selection.}
We adapt generalized binary splitting~\cite{hwang1972method} to identify individual key attributes among the remaining candidates. Tuple comparisons provide the relevance responses required by the group tests. With continued feedback, this phase ends when all candidates are classified or $d_\text{max}$ key attributes have been identified.

\paragraph{Phase 3: Regret Minimization.}
We adapt \textsc{UH-Random}~\cite{XCL19} to refine preferences over the retained attributes. The feasible utility region incorporates comparisons on displayed coordinates from Phases 1 and 2 and is refined through further interaction. With sufficient feedback, the method identifies a favorite tuple.

At any time before Phase 3, if the user stops the interaction, \textsc{AttributeSubset} returns $K$ tuples using the current candidate set $\mathcal{D}_\text{cand}$. At the extreme case with zero feedback, the candidate set could contain all $d$ attributes. If the user stops during interactive Phase 3, we return one recommendation using the current feasible utility region.

\subsection{Dimension Reduction}
\label{subsec:dimR}

Dimension reduction seeks a small candidate set containing every key attribute, using as few questions as possible. Starting with all attributes, Phase 1 eliminates irrelevant blocks and Phase 2 identifies individual key attributes, while maintaining $\lvert\mathcal{D}_\text{cand}\rvert\geq d_\text{int}$.

\paragraph{\textbf{User Interaction}} 
In Phases 1 and 2, each question presents $s$ tuples projected onto the attributes being tested. Only these attributes are displayed, so the number may vary between questions. Responses follow the partial-utility protocol in Section~\ref{sec:def} and update the candidate set $\mathcal{D}_\text{cand}$.

\subsubsection{\textbf{Phase 1: Coarse Elimination}}

We partition the $d$ attributes into $\lceil d/m\rceil$ disjoint blocks. For $i\in\{1,\ldots,\lceil d/m\rceil\}$, let $\mathcal{B}_i\subseteq\{D_1,\ldots,D_d\}$ denote the set of dimensions selected for the $i$-th Phase 1 question, defined by $\mathcal{B}_i=\{D_j:(i-1)m<j\leq\min(im,d)\}$. Each block contains at most $m$ attributes. The final block is displayed without padding when it contains fewer than $m$ attributes.

In each block, we:
\begin{enumerate}
  \item Construct the projected dataset $X_{\mathcal{B}_i}$ by restricting each tuple in $X$ to the dimensions in $\mathcal{B}_i$.
  \item Randomly select and present $s$ tuples from $X_{\mathcal{B}_i}$.
  \item If the user selects a tuple, retain all attributes in $\mathcal{B}_i$ as potentially key. If the user gives an irrelevant-attributes response, remove the entire block from $\mathcal{D}_\text{cand}$.
\end{enumerate}

Under the stated response model, every block containing a key attribute survives, while blocks containing only non-key attributes are eliminated. A complete Phase 1 uses at most $\lceil d/m\rceil$ questions and leaves $\lvert\mathcal{D}_\text{cand}\rvert\leq d_\text{int}m$: at most $d_\text{int}$ blocks survive, each containing at most $m$ attributes.

\begin{example}
    For $d=100$, $d_\text{int}=3$, and $m=7$, Phase 1 uses $15$ questions and leaves at most $21$ candidate attributes.
\end{example}

\subsubsection{\textbf{Phase 2: Fine Selection}}

Phase 2 applies generalized binary splitting~\cite{WGT}, a group testing algorithm to the remaining candidate attributes, to identify the key attributes. Because the true number of key attributes is unknown, which is a required input parameter by the group testing algorithm, we initialize its upper bound as $d_\text{left}=d_\text{max}=5$ and pass it as the input. Each identified key attribute decreases $d_\text{left}$ by one. Thus, $d_\text{left}$ bounds the number of keys still to be found; for example, after identifying $2$ keys, at most $3$ remain.

\begin{algorithm}[!t]
\caption{Phase 2: Fine Selection via Group Testing}
\label{alg:phase2}
\flushleft
\textbf{Input:} Candidate set $\mathcal{D}_\text{cand}$ of attributes (with $|\mathcal{D}_\text{cand}| \le d_\text{int} m$), upper bound $d_\text{max}$ on $d_\text{int}$ \newline
\textbf{Output:} Updated set $\mathcal{D}_\text{cand}$ containing all key attributes (larger than $d_\text{int}$ if the user quits earlier)

\begin{algorithmic}[1]

\State $d_\text{left} \gets d_\text{max}$, $\mathcal{D}_\text{cand}' \gets \emptyset$

\Statex {/*$\mathcal{D}_\text{cand}'$ contains identified key attributes, and $\mathcal{D}_\text{cand}$ contains unclassified attributes. The quantity $d_\text{left}=d_\text{max}-|\mathcal{D}_\text{cand}'|$ bounds the number of remaining key attributes.*/}
\Statex {/*Before every question, including individual tests and binary splitting, check whether the user stops. An unanswered question leaves its attributes unclassified.*/}

\While{$d_\text{left}>0$ and $\mathcal{D}_\text{cand}\neq \emptyset$}
    \If {user quits answering} break
    \EndIf
    \If{$|\mathcal{D}_\text{cand}| \leq 2d_\text{left} - 2$} \Comment{Base case}
        \For{each attribute $D_i \in \mathcal{D}_\text{cand}$}
            \If{the user stops}
                \State \Return $\mathcal{D}_\text{cand}\cup\mathcal{D}_\text{cand}'$
            \EndIf
            \State $\mathcal{D} \gets \{D_i\}$
            \State Present $s$ tuples from $X_\mathcal{D}$
            \If{the user stops without answering}
                \State \Return $\mathcal{D}_\text{cand}\cup\mathcal{D}_\text{cand}'$
            \EndIf
            \If{user selects a tuple}
                \State $\mathcal{D}_\text{cand}' \gets \mathcal{D}_\text{cand}' \cup \{D_i\}$, $d_\text{left} \gets d_\text{left} - 1$
            \EndIf
            \State $\mathcal{D}_\text{cand} \gets \mathcal{D}_\text{cand} \setminus \{D_i\}$
            \If{$d_\text{left}=0$} break
            \EndIf
        \EndFor
    \Else \Comment{Group testing}
        \State $l \gets |\mathcal{D}_\text{cand}| - d_\text{left} + 1$, $\alpha \gets \lfloor \log_2(l/d_\text{left}) \rfloor$
        \State $g \gets \min\{2^\alpha,m\}$
        \State Randomly sample $\mathcal{D} \subseteq \mathcal{D}_\text{cand}$, $|\mathcal{D}| = g$
        \State Present $s$ tuples from $X_\mathcal{D}$
        \If{the user stops without answering}
            \State \Return $\mathcal{D}_\text{cand}\cup\mathcal{D}_\text{cand}'$
        \EndIf
        \If{user selects a tuple}
            \State Apply balanced binary splitting to $\mathcal{D}$ until one key attribute $a^*$ is isolated or the user stops; let $\mathcal{D}_{-}$ contain the non-key attributes inferred from completed tests
            \State $\mathcal{D}_\text{cand}\gets\mathcal{D}_\text{cand}\setminus\mathcal{D}_{-}$
            \If{splitting stops before a key attribute is isolated}
                \State \Return $\mathcal{D}_\text{cand}\cup\mathcal{D}_\text{cand}'$
            \EndIf
            \State $\mathcal{D}_\text{cand}' \gets \mathcal{D}_\text{cand}' \cup \{a^*\}$, $d_\text{left} \gets d_\text{left} - 1$
            \State $\mathcal{D}_\text{cand} \gets \mathcal{D}_\text{cand} \setminus \{a^*\}$
        \Else
            \State $\mathcal{D}_\text{cand} \gets \mathcal{D}_\text{cand} \setminus \mathcal{D}$
        \EndIf
    \EndIf
\EndWhile
\State \Return $\mathcal{D}_\text{cand}\cup \mathcal{D}_\text{cand}'$
\end{algorithmic}
\end{algorithm}

Algorithm~\ref{alg:phase2} details the procedure. If $\lvert\mathcal{D}_\text{cand}\rvert\leq2d_\text{left}-2$, we test attributes individually. Otherwise, we test a random group of size $g=\min\{2^\alpha,m\}$, where $\alpha=\lfloor\log_2((\lvert\mathcal{D}_\text{cand}\rvert-d_\text{left}+1)/d_\text{left})\rfloor$. An irrelevant-attributes response eliminates the group. A tuple selection triggers balanced binary splitting to isolate one key attribute, eliminating any attributes identified as non-key along the way. Each question displays only the tested attributes. If the user stops, all unclassified attributes and previously identified keys are retained.

Phase 2 ends when all candidates are classified, $d_\text{max}$ key dimensions have been identified, or the user stops interacting. Under the default experimental settings, Phase 2 uses an average of $11$ additional questions.

\begin{example}
{\interlinepenalty=10000
We illustrate the first round of Phase~2 with $s=2$ and $m=6$. A response is \textit{positive} if the user selects a tuple and \textit{negative} if the user gives an irrelevant-attributes response. Each question displays only the tested attributes.\par}

Suppose the dataset has 100 attributes, with true key attributes $\{D_1,D_7,D_{13},D_{19},D_{25}\}$ ($d_\text{int}=5$), unknown to the algorithm. Starting from $\mathcal{D}_\text{cand}=\{D_1,\ldots,D_{100}\}$, Phase~1 with $m=6$ retains the first five blocks, leaving $\mathcal{D}_\text{cand}=\{D_1,\ldots,D_{30}\}$. Initially, $d_\text{left}=d_\text{max}=5$. Since $30>2d_\text{left}-2=8$, a group test is performed.

We compute $l=30-5+1=26$, $\alpha=\lfloor\log_2(26/5)\rfloor=2$, and $g=\min\{2^\alpha,m\}=4$. Suppose the sampled set is $\mathcal{D}=\{D_1,D_2,D_3,D_4\}$. We present two tuples from $X_\mathcal{D}$ on these four attributes. Since $\mathcal{D}$ contains $D_1$, the response is \textit{positive}, and binary splitting follows.

First, test $\{D_3,D_4\}$. The negative response identifies both attributes as non-key. The preceding positive group test therefore implies that $\{D_1,D_2\}$ contains a key attribute, without another question. Next, test $\{D_1\}$. The positive response identifies $D_1$ as key, while $D_2$ remains unclassified. Thus, the three questions display four, two, and one attributes, respectively.

Finally, add $D_1$ to $\mathcal{D}'_\text{cand}$. Here, $\mathcal{D}_{-}=\{D_3,D_4\}$. Update $\mathcal{D}_\text{cand}\gets\mathcal{D}_\text{cand}\setminus\{D_1,D_3,D_4\}$, leaving $30-2-1=27$ candidates, and decrement $d_\text{left}$ to 4. Repeat the procedure.
\end{example}

\begin{theorem}\label{theorem_q}
    Under error-free relevance responses, suppose at most $r$ key attributes remain among $N$ candidates. With continued feedback, standard generalized binary splitting requires at most $B(N,r)$ questions, where
    \[
    B(N,r)=\left\{\begin{array}{ll}
    0, & N=0\ \text{or}\ r=0,\\[2pt]
    N, & r\geq1,\ 1\leq N\leq 2r-2,\\[2pt]
    (\alpha+2)r+p-1, & r\geq1,\ N\geq 2r-1,
    \end{array}\right.
    \]
    where, in the third case, $l=N-r+1$, $\alpha=\left\lfloor\log_2(l/r)\right\rfloor$, and $p,\theta$ are defined by the unique decomposition
    \[
    l=2^\alpha r+2^\alpha p+\theta,\qquad 0\leq p<r,\qquad 0\leq\theta<2^\alpha.
    \]
    This bound adapts the standard generalized binary-splitting analysis~\cite{hwang1972method,Du1993CombinatorialGT}.
\end{theorem}

For FHDR, let $N=|\mathcal{D}_\text{cand}|$ at the start of Phase~2. A complete Phase~1 leaves $N\leq d_\text{int}m$ candidates. With $m=7$ and $d_\text{max}=5$, Phase~2 starts with $r=d_\text{left}=5$ and $N\leq35$. The analysis of the capped procedure covers every candidate count $0\leq N\leq35$ at this starting state and gives the same upper bound $B(N,5)$.

\begin{example}
If $d_\text{int}=3$, Phase~1 leaves at most 21 candidates. Since $B(21,5)=17$, Phase~2 requires at most 17 questions. If $d_\text{int}=5$, at most 35 candidates remain and $B(35,5)=21$. Thus, Phase~2 requires at most 21 questions under $m=7$, $d_\text{max}=5$, and $d_\text{int}\leq5$. These bounds concern completed attribute discovery and exclude Phase~3 questions.
\end{example}

A negative group test eliminates the tested attributes, whereas a positive test identifies one key attribute through binary splitting. Induction on the remaining problem size gives the standard bound. The detailed induction and the analysis of FHDR's capped procedure are provided in \iftoggle{TechnicalReport}{Section~\ref{appen:proof} of the \technicalreport}{Appendix~\ref{appen:proof}}.

The two-phase design reduces the display size before applying generalized binary splitting~\cite{WGT}. Applied directly to all $d$ attributes, its first group would contain $2^\alpha$ attributes, where $\alpha=\lfloor\log_2((d-d_\text{left}+1)/d_\text{left})\rfloor$. For $d=100$ and $d_\text{left}=5$, this gives $16$ displayed attributes. Phase 1 first reduces the candidate set to at most $d_\text{int}m$, keeping the initial Phase 2 group within the display limit, as formalized below.

\begin{theorem}\label{theorem_batch}
    The size of the batch for the first group test in Phase 2 would be no more than $m$.
\end{theorem}

By guaranteeing that every question involves at most $m$ attributes, this two-phase design preserves an efficient dimension reduction process, while ensuring a user-friendly interaction.

\FloatBarrier

\subsection{Regret Minimization}
\label{subsec:re_min}

After attribute discovery, FHDR applies an adapted \textsc{UH-Random} procedure~\cite{XCL19} over the retained attributes. With sufficient feedback, it identifies a favorite tuple. If the user stops before interactive Phase~3, \textsc{AttributeSubset} returns $K$ tuples using the current candidate attributes.

The adapted \textsc{UH-Random} procedure progressively refines the feasible utility region $\mathcal{R}$ over the retained attributes. Candidate tuples are retained while they can be optimal for a utility vector in this region. The procedure terminates when a favorite tuple is certified or the user stops. On early stopping, it returns a tuple maximizing utility under some $v\in\mathcal{R}$.

\begin{algorithm}[!t]
\caption{Overall Framework}
\label{alg:overall}
\flushleft
\textbf{Input:} A database $X$, desired output size $K$ under insufficient feedback, and an unknown utility vector $u$ (accessed through interactive queries)

\textbf{Output:}
\begin{enumerate}[label=(\arabic*)]
\item Sufficient feedback: $p^*\in\arg\max_{p\in X}u\cdot p$.
\item Stopping before Phase~3: $S\subseteq X$, $\lvert S\rvert=K$.
\end{enumerate}

\begin{algorithmic}[1]
\State $\mathcal{D}_{\mathrm{cand}}\gets\{D_1,\ldots,D_d\}$, $\mathcal{C}\gets\emptyset$ \Comment{comparison constraints}
\State Select $m$ and $s$ \Comment{internal parameters (default: $m=7$, $s=2$)}

\While{attribute discovery is incomplete and the user continues answering}
    \State Select $\mathcal{D}$ according to the current Phase~1 or Phase~2 test
    \State Randomly sample and display $s$ tuples projected onto $\mathcal{D}$, and receive feedback
    \If{the user stops without answering}
        \State \textbf{break}
    \EndIf
    \State Update $\mathcal{D}_{\mathrm{cand}}$ according to the discovery rules
    \State Add $a_{\mathcal{D}}(p,q)$ to $\mathcal{C}$ for each elicited strict comparison
\EndWhile

\If{the user stops before entering Phase~3}
    \State \Return $\textsc{AttributeSubset}(X,\mathcal{D}_{\mathrm{cand}},K)$
\EndIf

\State Initialize the feasible utility region $\mathcal{R}$ over the retained attributes using $\mathcal{C}$
\Repeat
    \If{a tuple $p^*$ is certified as a favorite for every utility vector in $\mathcal{R}$}
        \State \Return $p^*$
    \EndIf
    \If{the user stops}
        \State Select $v\in\mathcal{R}$
        \State \Return any $\hat p\in\arg\max_{p\in X}v\cdot p_{\mathcal{D}_{\mathrm{cand}}}$
    \EndIf
    \State Perform the next comparison prescribed by the adapted \textsc{UH-Random} procedure and update $\mathcal{R}$
\Until{a recommendation is returned}
\end{algorithmic}
\end{algorithm}

In Phases 1 and 2, relevance responses update the candidate attribute set. Tuple comparisons can also constrain relative utility weights. For tuples $p,q$ compared on the displayed attribute set $\mathcal{D}$, define the difference vector $a_{\mathcal{D}}(p,q)\in\mathbb{R}^d$ by
\[
a_{\mathcal{D}}(p,q)[i]=
\begin{cases}
p[i]-q[i],&D_i\in\mathcal{D},\\
0,&D_i\notin\mathcal{D}.
\end{cases}
\]

A strict preference for $p$ over $q$ on the displayed attributes gives $f_{\mathcal{D}}(p)>f_{\mathcal{D}}(q)$, equivalently $u\cdot a_{\mathcal{D}}(p,q)>0$. To maintain a closed polytope represented by its vertices, we use the conservative closed halfspace
\[
H_{p,q,\mathcal{D}}=\{v\in\mathbb{R}^d:v\cdot a_{\mathcal{D}}(p,q)\geq0\}.
\]
Its boundary is $v\cdot a_{\mathcal{D}}(p,q)=0$, and the true utility vector remains feasible. Geometrically, the strict preference implies $\theta(u,a_{\mathcal{D}}(p,q))<90^\circ$. A tuple certified as a favorite throughout the maintained region is therefore also a favorite under $u$.

At least two key attributes must be displayed for a comparison to constrain their relative weights. If no key attribute is displayed, the user gives an irrelevant-attributes response. If exactly one key attribute $i$ is displayed, the inequality reduces to $u[i](p[i]-q[i])>0$. Since $u[i]>0$, this is equivalent to $p[i]>q[i]$ and provides no additional information about relative utility weights.

We retain these displayed-coordinate difference vectors together with their original attribute indices. When initializing \textsc{UH-Random}, we restrict each vector to the retained candidate attributes and use the resulting constraints to initialize the feasible utility region. Under the stated response model, eliminated attributes have zero utility weight, so removing their coordinates preserves each constraint.

\subsection{Summary}
\label{subsec:sum}

FHDR uses coarse elimination and generalized binary splitting to reduce the candidate attribute space without upfront attribute selection, then produces recommendations according to the available feedback. With sufficient feedback, FHDR identifies a favorite tuple. If interaction ends before Phase~3, \textsc{AttributeSubset} returns $K$ recommendations; if it ends during Phase~3, FHDR returns one recommendation based on the current feasible utility region. The pseudocode for the overall algorithm is presented in Algorithm \ref{alg:overall}.

\FloatBarrier

%% file: attributesubset.tex
\section{Single-Round Algorithm}
\label{sec:attsub}

We propose \textsc{AttributeSubset}, a single-round algorithm for Problem~2~[\ref{p2}]. It selects $K$ tuples from $X$ using the current candidate attributes $\mathcal{D}_\text{cand}$, without further feedback.

\paragraph{\textbf{Motivation}}
When interaction ends before Phase 3, the candidate set may be reduced but still moderately large (e.g., about $30$ attributes). Single-round methods such as \textsc{Sphere}~\cite{XW18} and \textsc{CoresetHS}~\cite{kumar2018faster} claim to be applicable to any dimensionality, but their performance degrades severely in high-dimensional settings. Take \textsc{Sphere} for example, the theoretical bounds $O(n e^{O(\sqrt{d \log n})} + nk^3d)$~\cite{Gar95} and $O(nkd^2 2^d + nk^3d)$~\cite{SW92} grow sub-exponentially with $d$. Its requirement of $k\geq d$ results in overly large recommendation sets (e.g., $k \ge 100$ when $d = 100$), which are unsuitable for real-world use. \textsc{Sphere} also optimizes worst-case regret over the full utility space without exploiting sparse preferences.

\paragraph{\textbf{Inner Runs}}
To address these limitations, we propose the \textsc{AttributeSubset} algorithm, which consists of multiple inner runs. Each inner run $i$ independently samples a uniformly random $w$-element subset of $\mathcal{D}_\text{cand}$. We compute the skyline of the projected dataset using \textsc{BSkyTree}~\cite{bskytree} and apply \textsc{Sphere} with inner output budget $k>w$, obtaining $S_i$. Attributes are distinct within a sample and may recur across runs. We use $w=6$ as an experimental default, balancing execution time and regret in the parameter study (Figure~\ref{fig:w}). We choose $k=w+1=7$, the smallest integer satisfying $k>w$. \textsc{BSkyTree} uses balanced pivot-based partitioning to prune dominated regions and avoid comparisons between incomparable regions.

\paragraph{\textbf{Aggregation}}
Let $U=\bigcup_{i=1}^{N}S_i$. We stop when $\lvert U\rvert\geq K$ or $max\_iter$ inner runs have completed. If $\lvert U\rvert<K$, we add $K-\lvert U\rvert$ random tuples from $X\setminus U$; if $\lvert U\rvert>K$, we uniformly sample $K$ distinct tuples from $U$. Denote $S$ as the resulting set. The pseudocode is presented in Algorithm~\ref{alg:att}.

\paragraph{\textbf{Small Candidate Sets}}
If $\lvert\mathcal{D}_\text{cand}\rvert\leq w$, we use all candidate attributes in a single skyline--\textsc{Sphere} run with output budget $K$.

\begin{algorithm}[!t]
\caption{\textsc{AttributeSubset}}
\label{alg:att}
\flushleft
\textbf{Input:} A database $X$, a candidate set $\mathcal{D}_\text{cand}$ and the final output size $K$ \\
\textbf{Output:} A subset $S\subseteq X$ with low regret ratio

\begin{algorithmic}[1]

\State $S \gets \emptyset$

\Statex /*default value: $w=6, k = 7, max\_iter = 50$*/

\State Select some $w \geq d_\text{max}$, $k>w$, $max\_iter$ \Comment{internal parameters}

\If {$w \geq |\mathcal{D}_\text{cand}|$}
    \State Construct a new dataset $X_{\mathcal{D}_\text{cand}}$ restricted to dimensions in $\mathcal{D}_\text{cand}$
    \State $X_{\mathcal{D}_\text{cand}} = $ \textsc{Skyline}$(X_{\mathcal{D}_\text{cand}})$, $S = $ \textsc{Sphere}$(X_{\mathcal{D}_\text{cand}}, K)$
    \State return S
\EndIf

\State $iter \gets 0$
\While {$|S|<K$ and $iter < max\_iter$}
    \State Independently sample $w$ distinct dimensions uniformly from $\mathcal{D}_\text{cand}$
    \State Construct a new dataset $X_{\mathcal{D}_w}$ restricted to the $w$ dimensions
    \State $X_{\mathcal{D}_w} = $ \textsc{Skyline}$(X_{\mathcal{D}_w})$, $S_i = $ \textsc{Sphere}$(X_{\mathcal{D}_w}, k)$
    \State Convert points in $S_i$ back to the original dimension $d$
    \State $S := S \cup S_i$
    \State $iter \gets iter+1$
\EndWhile

\If {$|S| < K$}
    \State $S := S \cup \{\text{$K-|S|$ random points from $X\setminus S$} \}$
\EndIf

\If {$|S| > K$}
    \State $S := $ $K$ random points from $S$
\EndIf

\State return $S$
\end{algorithmic}
\end{algorithm}

\paragraph{\textbf{Regret Transfer and Trimming}}
The following theorem transfers an inner output's regret bound to the original dataset when all key attributes are sampled.

\begin{theorem}\label{theorem_att}
    Fix $u$, and suppose the sampled set $\mathcal{D}_w$ contains all its key attributes. Let $S_i\subseteq X$ denote the original tuples corresponding to the inner output. If its projected output has regret ratio at most $\varepsilon$ for every nonnegative unit-$L_1$ utility vector over $\mathcal{D}_w$, then $\mathrm{rr}_X(S_i,u)\leq\varepsilon$.
\end{theorem}

Omitted attributes have zero weight, so original and projected utilities coincide. See \iftoggle{TechnicalReport}{Section~\ref{appen:proof} of the \technicalreport}{Appendix~\ref{appen:proof}}.

The union $U$ inherits any regret bound satisfied by an inner output $S_i$, and adding tuples preserves this bound. To obtain exactly $K$ tuples, we uniformly trim an oversized union. Since the last run adds at most $k$ tuples, $|U|\leq K+k-1$. For a fixed utility vector $u$, trimming retains a specified utility-maximizing tuple in $U$ with probability $K/|U|\geq K/(K+k-1)$, preserving the union's regret bound with at least this probability. See \iftoggle{TechnicalReport}{Section~\ref{appen:proof} of the \technicalreport}{Appendix~\ref{appen:proof}} for a proof.

\paragraph{\textbf{Coverage Analysis}}
Fix a user's key-attribute set $\mathcal{H}\subseteq\mathcal{D}_\text{cand}$. Let $D=|\mathcal{D}_\text{cand}|$ and $h=|\mathcal{H}|=d_\text{int}$. Consider the repeated-sampling branch, where $1\leq h\leq w<D$. Let $P_\text{cover}$ be the probability that a sample contains $\mathcal{H}$.
\begin{lemma}\label{lemma_p}
    Under uniform sampling,
    \[
    P_\text{cover}=\frac{\binom{D-h}{w-h}}{\binom Dw}
    =\frac{\binom wh}{\binom Dh}
    \geq\left(\frac{w-h+1}{D}\right)^h.
    \]
\end{lemma}

Exactly $\binom{D-h}{w-h}$ samples contain the fixed key-attribute set; see \iftoggle{TechnicalReport}{Section~\ref{appen:proof} of the \technicalreport}{Appendix~\ref{appen:proof}}.

For a fixed positive integer $t$, let $C_t$ denote the event that at least one of $t$ independent attribute samples contains every key attribute. Then
\[
\Pr(C_t)=1-(1-P_\text{cover})^t.
\]
For a target coverage confidence $\gamma\in(0,1)$, the minimum prescribed number of independent samples is
\[
t_\gamma=\left\lceil\frac{\ln(1-\gamma)}{\ln(1-P_\text{cover})}\right\rceil,
\qquad 0<P_\text{cover}<1.
\]
These expressions quantify coverage for a prescribed number of independent samples. \textsc{AttributeSubset} stops according to the accumulated union size and the iteration cap.

In the reported Car example, $12$ interaction questions leave $D=12$ candidate attributes, with $h=3$ and $w=6$. \textsc{AttributeSubset} completes $33$ inner runs in approximately $0.1$~s. For a prescribed sequence of $33$ independent samples, the coverage probability is $1-(10/11)^{33}\approx95.7\%$.

Overall, the \textsc{AttributeSubset} algorithm effectively bridges the gap between the general applicability of \textsc{Sphere} and the practical constraints of high-dimensional regret minimization with sparse user preferences, introducing both efficiency and accuracy.

\FloatBarrier

%% file: experiment.tex
\section{Experiments}
\label{sec:eval}

We conducted experiments with an Apple M2 Pro chip and 32GB of RAM. All programs were implemented in C/C++.

\paragraph{\textbf{Datasets. }}
We generate synthetic datasets with independent, uniformly distributed attributes using the skyline data generator~\cite{borzsony2001skyline}.

We also evaluate five high-dimensional real datasets: NBA~\cite{basketballReference}, with $4{,}790$ player-season tuples and $104$ attributes; Car~\cite{epaAutomotiveTrends}, with $5{,}500$ vehicles and $57$ attributes; House~\cite{house_austin}, with $15{,}171$ properties and $33$ attributes; Energy~\cite{large-scale_wave_energy_farm_882}, with $36{,}043$ wave-energy farm configurations and $149$ attributes; and Activity~\cite{pamap2_physical_activity_monitoring_231}, with $2{,}872{,}533$ processed activity profiles and $54$ attributes. Detailed statistics appear in \iftoggle{TechnicalReport}{Section~\ref{appen:setting} of the \technicalreport}{Appendix~\ref{appen:setting}}.

We encoded missing values using the minimum observed value of the corresponding attribute. We removed constant attributes and normalized the remaining attributes to $[0,1]$, reversing those for which smaller values are preferred, as described in Section~\ref{sec:def}. All reported dimensionalities refer to the retained attributes.

\paragraph{\textbf{Algorithms. }}
We compare \textsc{FHDR} with two baselines: (1) \textsc{UtilityApprox}~\cite{NL12}, an interactive method applicable to high-dimensional data; and (2) \textsc{Sphere-Adapt}, an adaptation of the single-round \textsc{Sphere} algorithm~\cite{XW18}. For the set-output experiments, we adapt \textsc{UtilityApprox} to return the $K$ highest-scoring tuples under its estimated utility vector. To ensure fairness, \textsc{Sphere-Adapt} receives the candidate attribute set obtained by FHDR’s dimension reduction process and runs \textsc{Sphere} on the corresponding projected data. This comparison evaluates \textsc{AttributeSubset} against \textsc{Sphere} within the same discovered attribute space.

Table~\ref{tab:capabilities} summarizes the practical limitations of additional methods. In the tested high-dimensional settings, \textsc{UH-Random}/\textsc{UH-Simplex} and \textsc{HD-PI} encounter convex-hull computation failures, while \textsc{Verify-Point} encounters Qhull errors and excessive memory use. \textsc{Breakpoint} requires a counter repair on the large synthetic input and incurs substantial interaction and output costs. Detailed reproduction results are provided in \iftoggle{TechnicalReport}{Section~\ref{appen:capabilities} of the \technicalreport}{Appendix~\ref{appen:capabilities}}. We retain \textsc{UtilityApprox} and \textsc{Sphere-Adapt} for quantitative comparison and evaluate \textsc{UH-Random}’s upfront attribute-selection mechanism in the user study in Section~\ref{sec:user_study}.

\paragraph{\textbf{Parameter Setting. }}
We evaluate each algorithm by varying five key parameters: (1) dataset size $n$, (2) dimensionality $d$, (3) number of key attributes $d_\text{int}$, (4) number of answered questions $q$, and (5) output size $K$ for Problem~2~[\ref{p2}].

Unless otherwise specified, we use $d_\text{int}=3$~\cite{GG96, GA12, WX22, SC05, TK03, DMW12}, question size $s=2$~\cite{NL12}, question budget $q=15$ (see [\ref{q}] for justification), output size $K=30$~\cite{XW18}, and target regret ratio $\varepsilon=0$ for \textsc{UH-Random}. For synthetic datasets, we set $n=10^5$~\cite{NL12, WWX21, XCL19, XW18} and $d=100$.

We study two cases: complete interaction and limited feedback, with $q=15$ or $q=0$. We also vary the maximum number of displayed attributes $m$ and the number sampled per \textsc{AttributeSubset} inner run $w$ on real datasets across different $d_\text{int}$. Based on these results, we use $m=7$ and $w=6$; further details appear in \iftoggle{TechnicalReport}{Section~\ref{appen:setting} of the \technicalreport}{Appendix~\ref{appen:setting}}.

\paragraph{\textbf{Performance Measurement. }}
We evaluate the algorithms using four metrics: (1) \textit{execution time}, the end-to-end time (excluding user response time) for the final output; (2) \textit{regret ratio}, the relative utility loss between the best returned tuple and the global optimum under the ground-truth utility function; (3) \textit{outperformance rate}, the fraction of repeated runs in which our algorithm achieves the lowest regret ratio, with ties broken by execution time; and (4) \textit{number of questions}, which measures user effort.

Each experiment is repeated $100$ times with independently generated utility functions, and we report the averages.

\subsection{Results on Synthetic Datasets}
We evaluate sufficient- and limited-feedback settings separately. All experimental curves labeled \textsc{Sphere} correspond to \textsc{Sphere-Adapt}, as defined above.

\subsubsection{Problem 1}[\ref{p1}]
\begin{figure*}[ht]
  \centering

  \begin{minipage}[t]{0.32\textwidth}
    \centering
    \renewcommand{\arraystretch}{1.2}
    \includegraphics[width=\linewidth]{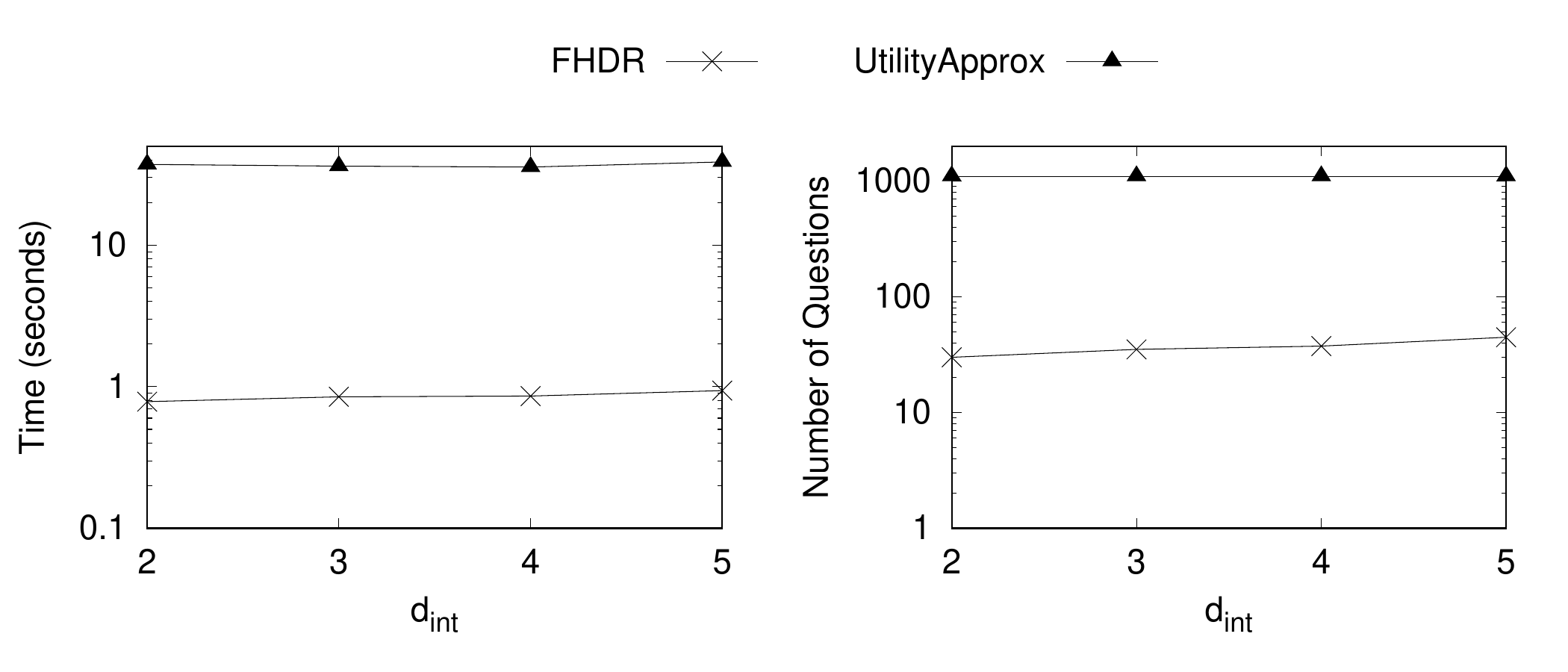}
    \caption{P1: Vary $d_\text{int}$}
    \label{fig:p1_d_int}
  \end{minipage}
  \hfill
  \begin{minipage}[t]{0.32\textwidth}
    \centering
    \renewcommand{\arraystretch}{1.2}
    \includegraphics[width=\linewidth]{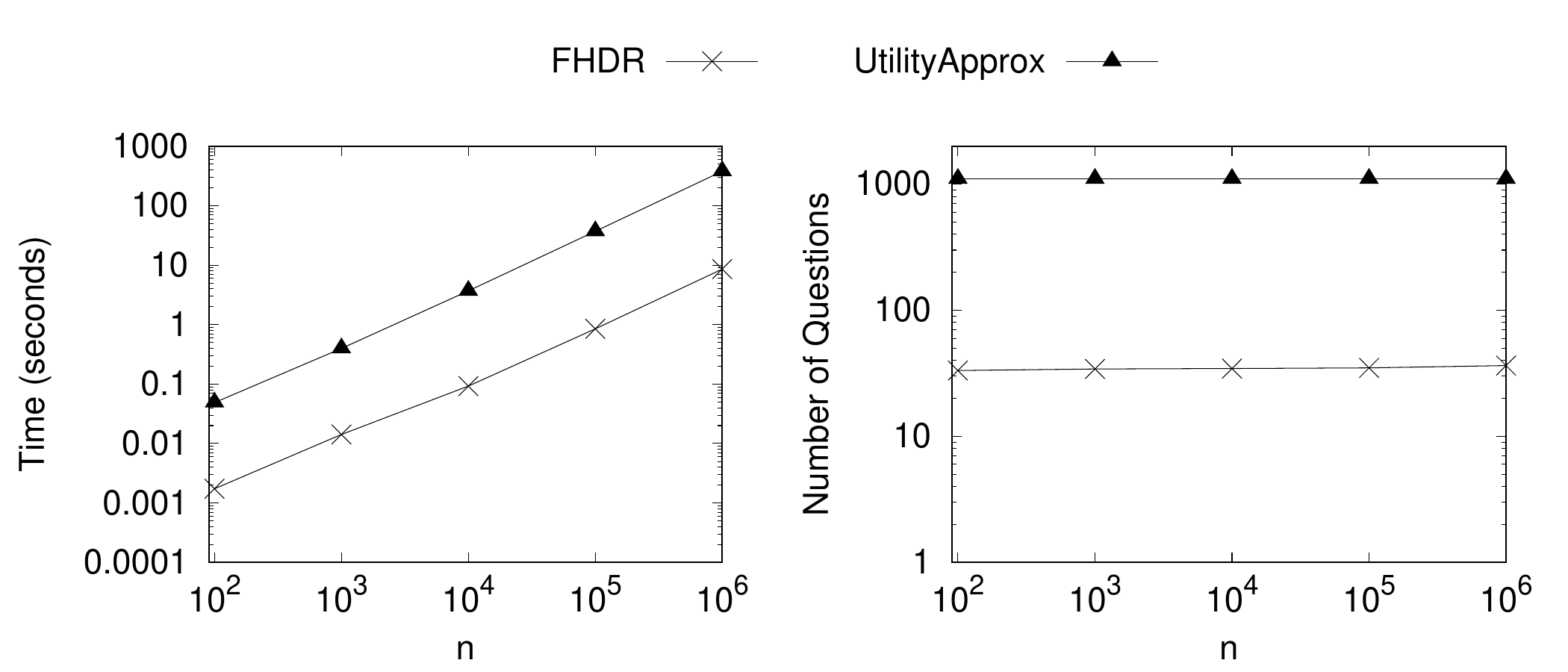}
    \caption{P1: Vary $n$}
    \label{fig:p1_n}
  \end{minipage}
  \hfill
  \begin{minipage}[t]{0.32\textwidth}
    \centering
    \centering
    \includegraphics[width=\linewidth]{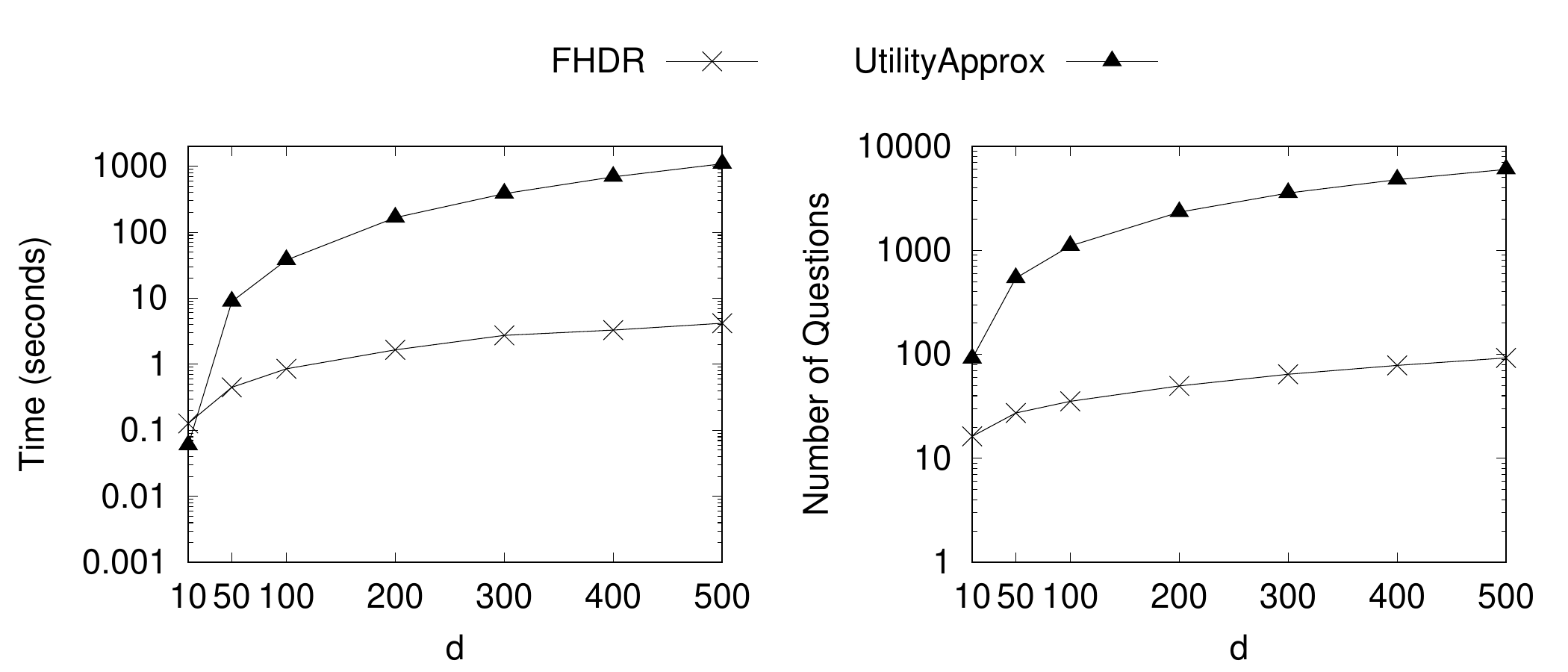}
    \caption{P1: Vary $d$}
    \label{fig:p1_d}
  \end{minipage}

\end{figure*}

We first evaluate favorite-tuple retrieval under sufficient feedback, reporting the number of questions and execution time. FHDR terminates when its recommendation is certified as a favorite tuple. For \textsc{UtilityApprox}, we set $\varepsilon=0$ and terminate when $\sum_{i=1}^{d}(U_i-L_i)<10^{-9}$ or the $10{,}000$-question cap is reached, where $L_i$ and $U_i$ are its utility-coefficient bounds. Its reported costs correspond to this numerical convergence criterion. We compare with \textsc{UtilityApprox} here and evaluate the single-round \textsc{Sphere-Adapt} in the limited-feedback setting.

We vary the number of key attributes $d_\text{int}$ from $2$ to $5$, following the preference-sparsity setting motivated above. As shown in Figure~\ref{fig:p1_d_int}, the number of questions required by our algorithm increases moderately from $29$ to $42$, whereas \textsc{UtilityApprox} consistently requires approximately $1100$ questions, making it impractical for real-world use. The execution time remains low, increasing only from $0.02$~s to $0.05$~s, while \textsc{UtilityApprox} requires approximately $37$~s. The nearly constant behavior of \textsc{UtilityApprox} follows from its weakly truthful design: it refines the hidden utility function using synthetic tuples rather than actual data points. Consequently, for a fixed dimensionality $d$, the number of questions is largely independent of the input dataset. With a fixed number of tuples, its execution time therefore also remains nearly constant.

Figures~\ref{fig:p1_n} and~\ref{fig:p1_d} evaluate scalability with respect to the number of tuples $n$ and the dimensionality $d$. As shown in Figure~\ref{fig:p1_n}, our algorithm scales well with $n$, completing within a few seconds even when $n=10^6$. In contrast, the execution time of \textsc{UtilityApprox} grows approximately linearly with $n$, reaching $378$~s at $n=10^6$. For fixed $d$, our algorithm requires approximately $34$ questions regardless of $n$, whereas \textsc{UtilityApprox} requires about $1100$. As $d$ increases from $10$ to $500$, FHDR's question count grows much more slowly than \textsc{UtilityApprox}'s. By comparison, \textsc{UtilityApprox} requires $91$ questions at $d=10$ and $5989$ questions at $d=500$. The execution-time results exhibit a similar trend.

\subsubsection{Problem 2}[\ref{p2}]

\begin{figure}[ht]
    \centering
    \begin{minipage}{\linewidth}
        \centering
        \includegraphics[width=\linewidth]{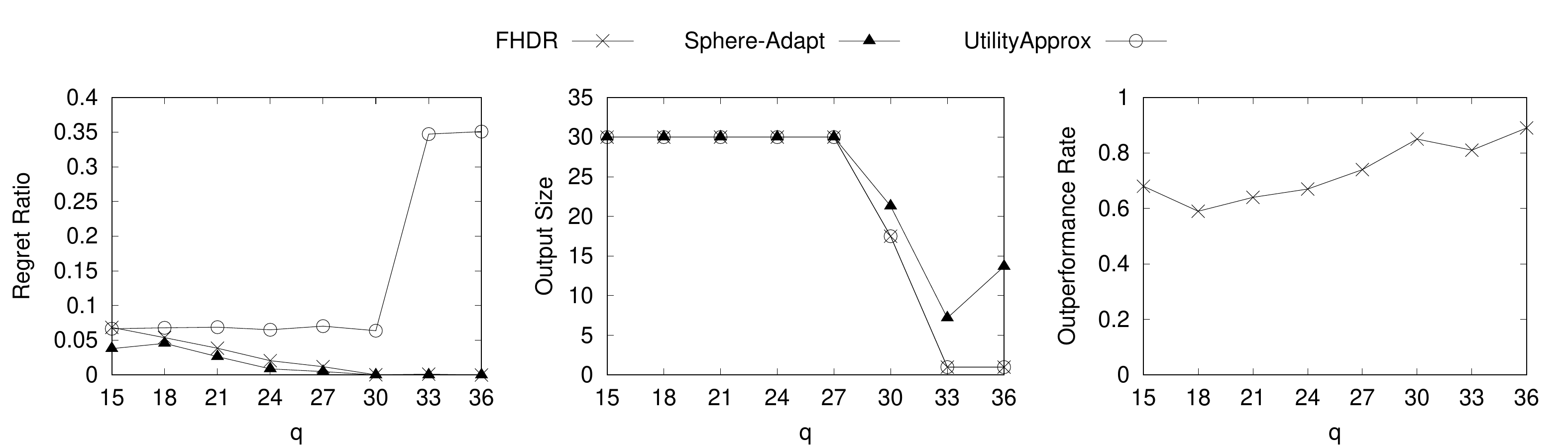}
        \vspace{-1em}
        \caption{P2: Vary $q$}
        \label{fig:p2_q}
    \end{minipage}

    \begin{minipage}{\linewidth}
        \centering
        \includegraphics[width=\linewidth]{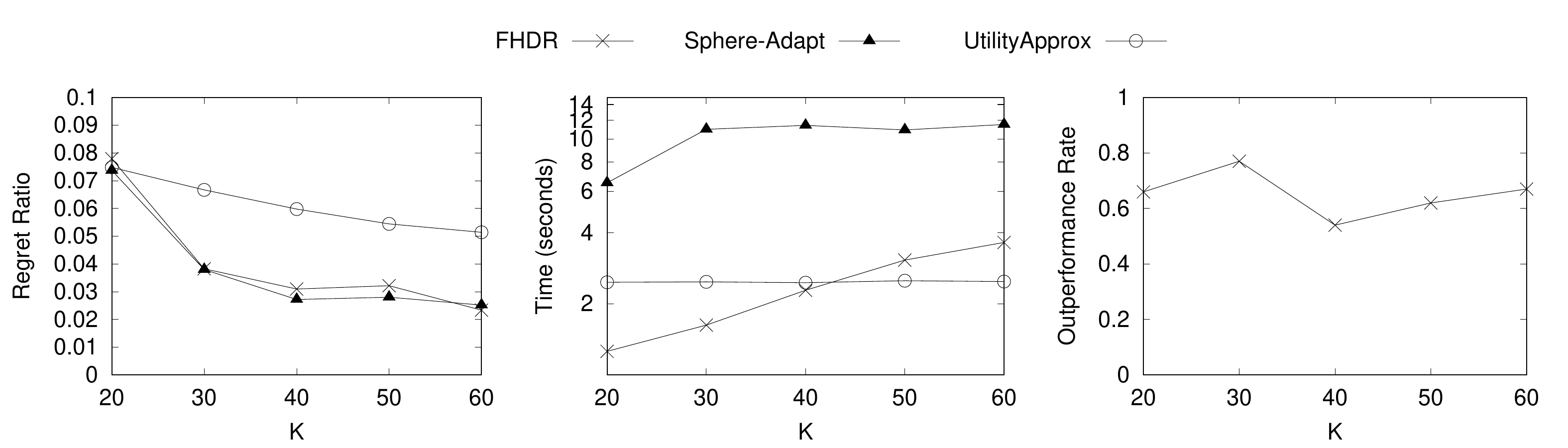}
        \vspace{-1em}
        \caption{P2: Vary $K$}
        \label{fig:p2_K}
    \end{minipage}
    
    \begin{minipage}{\linewidth}
        \centering
        \includegraphics[width=\linewidth]{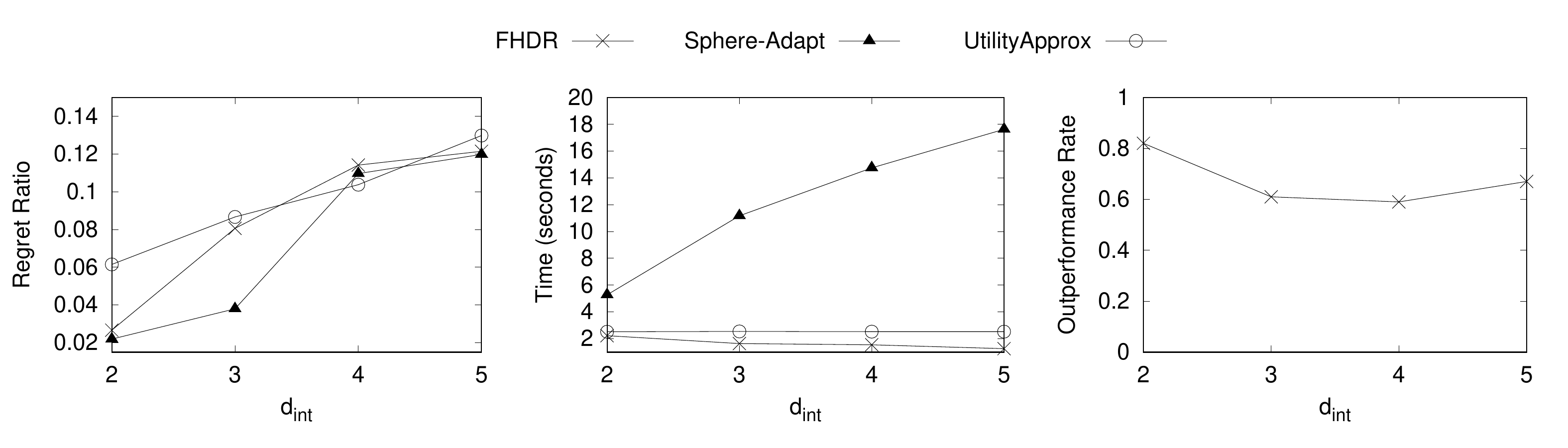}
        \vspace{-1em}
        \caption{P2: Vary $d_\text{int}$}
        \label{fig:p2_d_int}
    \end{minipage}

    \begin{minipage}{\linewidth}
        \centering
        \includegraphics[width=\linewidth]{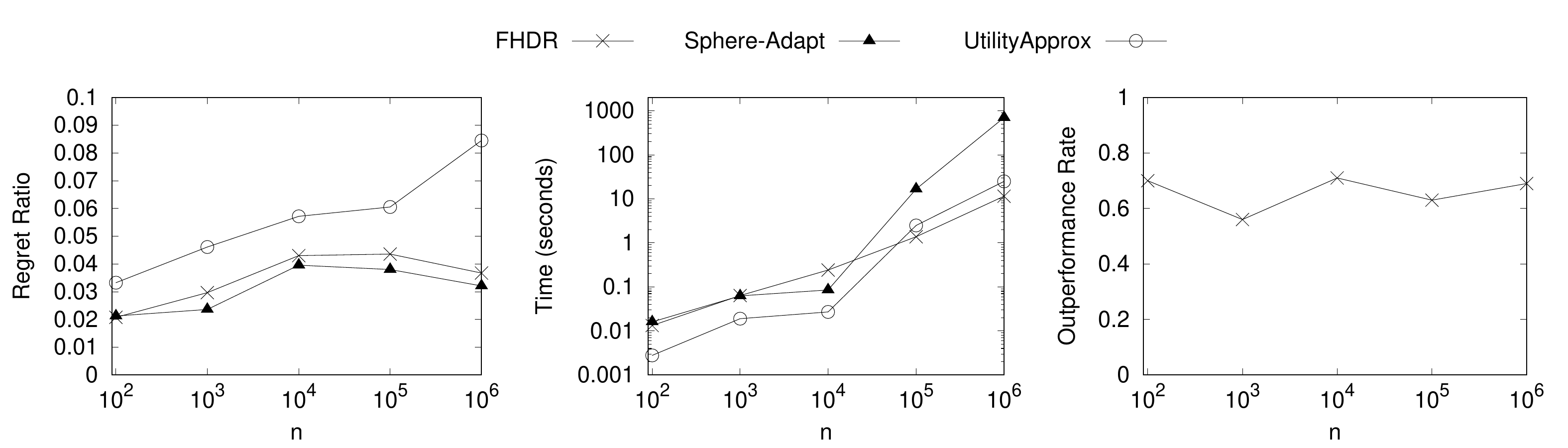}
        \vspace{-1em}
        \caption{P2: Vary $n$}
        \label{fig:p2_n}
    \end{minipage}
    
    \begin{minipage}{\linewidth}
        \centering
        \includegraphics[width=\linewidth]{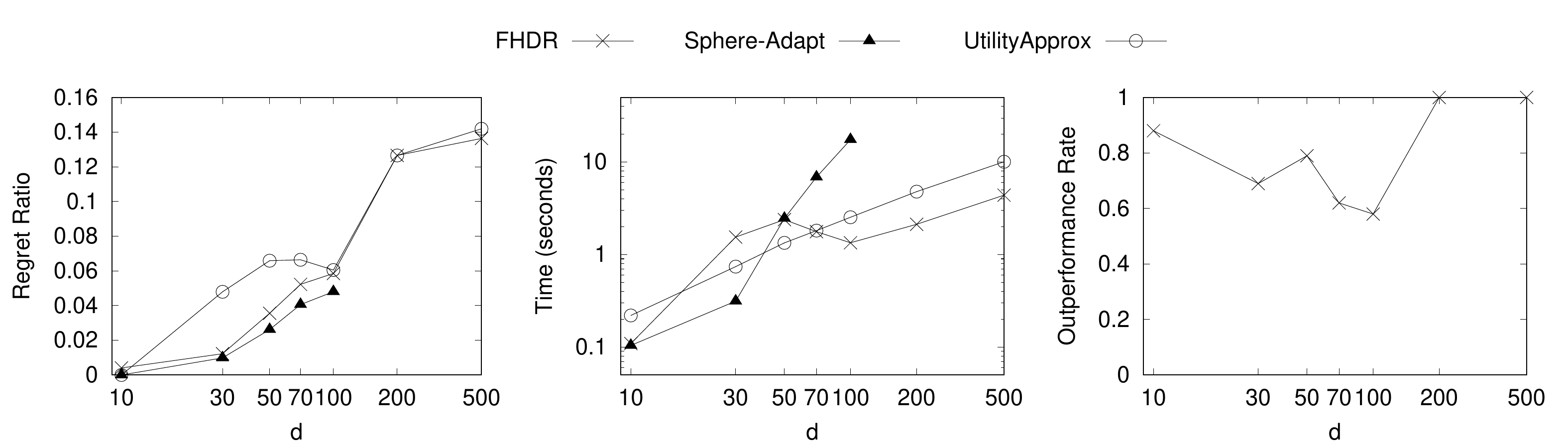}
        \vspace{-1em}
        \caption{P2: Vary $d$}
        \label{fig:p2_d}
    \end{minipage}
    \begin{minipage}{\linewidth}
        \centering
        \includegraphics[width=\linewidth]{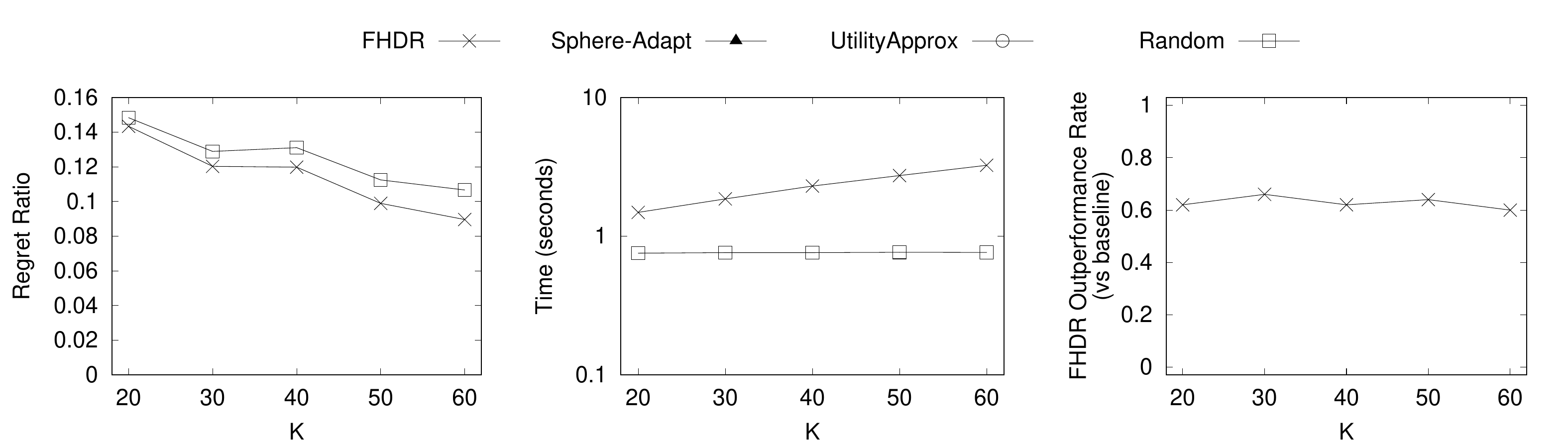}
        \vspace{-1em}
        \caption{$P2^{*}$: Vary $K$}
        \label{fig:p2star_synthetic}
    \end{minipage}
\end{figure}

We next evaluate recommendation quality under fixed question budgets.

\label{q}
Figure~\ref{fig:p2_q} varies $q$ from $15$ to $36$ in increments of $3$. Under the default settings, $15=\lceil100/7\rceil$ questions complete Phase 1 and leave at most $d_\text{int}m=21\leq K=30$ candidate attributes, allowing comparison with \textsc{Sphere-Adapt} under the same output budget.

Figure~\ref{fig:p2_q} shows FHDR reaching zero regret with sufficient Phase~3 feedback. Based on aforementioned algorithm design, if the user stops before Phase~3, FHDR returns a recommendation set; if interaction continues into Phase~3, FHDR returns one tuple and progressively refines the recommendation. Before FHDR enters the interactive Phase~3 branch, all methods return $K=30$ tuples. Afterward, FHDR and \textsc{UtilityApprox} return one tuple, while \textsc{Sphere-Adapt} uses the output budget $K=sq'$, where $q'$ counts FHDR’s additional Phase~3 questions and $s$ is the number of tuples displayed per question. Following the tuple-presentation accounting in~\cite{XCL19}, this budget gives the single-round method a corresponding allowance for FHDR’s additional interaction. \textsc{UtilityApprox}'s regret increase coincides with its output shrinking from $K=30$ tuples to one, as shown in the output-size panel. Returning fewer tuples can increase regret despite additional feedback; the curve therefore reflects changes in both the question budget and output size. Our method consistently achieves a lower regret ratio compared with \textsc{UtilityApprox}. While \textsc{Sphere-Adapt} attains similar regret ratios, our method performs better in over $60\%$ of cases.

We vary the output size $K$ in Figure~\ref{fig:p2_K}. FHDR achieves lower regret ratios than \textsc{UtilityApprox}. Its execution time increases from $1$ to $3$ seconds, compared with $\sim11$ seconds for \textsc{Sphere-Adapt}. Although slightly slower than \textsc{UtilityApprox}, it provides substantially better result quality and achieves the lowest regret ratio in $\sim60\%$ of the runs.

Figure~\ref{fig:p2_d_int} shows the effect of varying $d_\text{int}$, which reflects increasing preference complexity. For all three algorithms, the regret ratio grows approximately linearly as $d_\text{int}$ increases from $2$ to $5$, with lower regret observed at smaller values. Execution times remain consistent with earlier results, although our algorithm becomes slightly faster as $d_\text{int}$ increases. With a fixed question budget, a larger $d_\text{int}$ produces a larger candidate set $|\mathcal{D}_\text{cand}|$, reducing overlap among the inner output sets $S_i$. The union can therefore reach the fixed output size $K$ in fewer inner runs, reducing execution time.

Figures~\ref{fig:p2_n} and~\ref{fig:p2_d} evaluate scalability with respect to the number of tuples $n$ and dimensionality $d$. As $n$ increases, our algorithm maintains a regret ratio below $0.045$, substantially lower than \textsc{UtilityApprox}, whose regret increases noticeably. FHDR's runtime grows approximately linearly with $n$, whereas \textsc{Sphere-Adapt} scales less favorably. Section~\ref{sec:attsub} discusses \textsc{Sphere}'s dimension-dependent cost.

As shown in Figure~\ref{fig:p2_d}, our algorithm also consistently outperforms \textsc{UtilityApprox} in regret ratio and remains close to \textsc{Sphere-Adapt}. Its runtime stays within a few seconds, first peaking at $d=50$, then decreasing slightly before rising again. The temporary decrease occurs because a larger $d$ produces a larger candidate set $|\mathcal{D}_\text{cand}|$ under a fixed question budget, reducing overlap among the inner output sets and hence the number of \textsc{Sphere} runs within \textsc{AttributeSubset}. At larger $d$, dataset-loading costs begin to dominate. \textsc{Sphere-Adapt} becomes infeasible beyond $d=100$ under the default $K$ and $q$ settings, whereas our method continues to return valid results in approximately one second, demonstrating the robustness in extremely high-dimensional settings.

\paragraph*{Zero-Feedback Evaluation ($P2^{*}$)}
We also evaluate the special case of zero-feedback in Problem~2, denoted by $P2^{*}$, in which the user answers no questions ($q=0$). FHDR directly invokes \textsc{AttributeSubset} on all original attributes, without attribute elimination or elicited comparison constraints. We compare our method with \textsc{Random}, a baseline which samples $K$ distinct tuples uniformly without replacement from the entire dataset.

Figure~\ref{fig:p2star_synthetic} shows that FHDR achieves lower mean regret than \textsc{Random} at every tested $K$. As $K$ increases from $20$ to $60$, FHDR’s mean regret decreases from $0.1434$ to $0.0897$, compared with $0.1483$ to $0.1067$ for \textsc{Random}. FHDR achieves lower or equal regret in $60$--$66\%$ of runs.

\subsection{Results on Real Datasets}
We report NBA results for sufficient and limited feedback; results for the other real datasets appear in \iftoggle{TechnicalReport}{Section~\ref{appen:real} of the \technicalreport}{Appendix~\ref{appen:real}}.

\begin{figure}[ht]
    \centering
    \begin{subfigure}{\linewidth}
        \centering
        \includegraphics[width=0.67\linewidth]{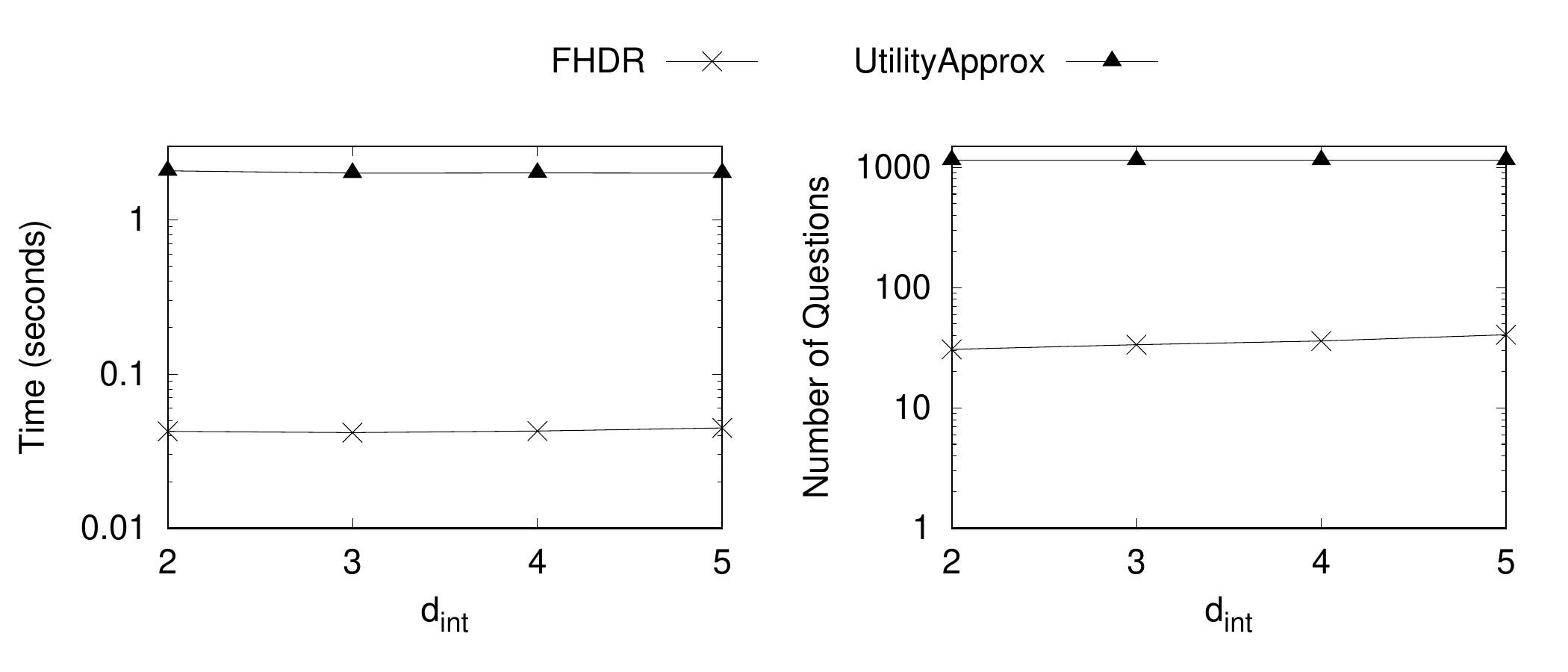}
        \vspace{-0.5em}
        \caption{P1: Vary $d_{\text{int}}$}
        \label{fig:nba_a}
      \end{subfigure}

    \begin{subfigure}{\linewidth}
        \centering
        \includegraphics[width=\linewidth]{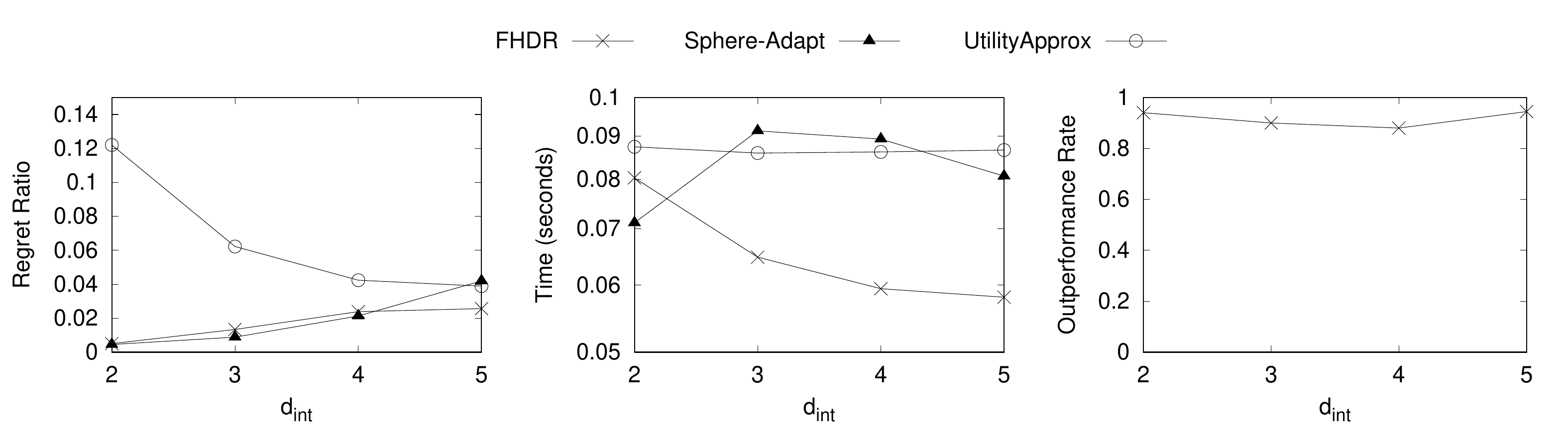}
        \vspace{-1.5em}
        \caption{P2: Vary $d_{\text{int}}$}
        \label{fig:nba_b}
      \end{subfigure}
    
    \begin{subfigure}{\linewidth}
        \centering
        \includegraphics[width=\linewidth]{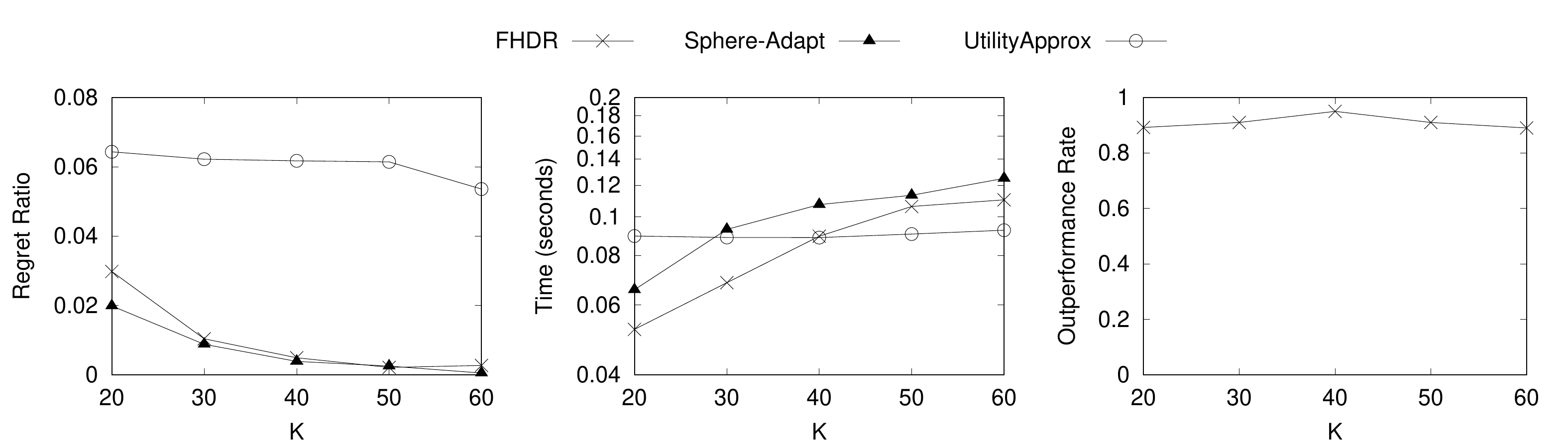}
        \vspace{-1.5em}
        \caption{P2: Vary $K$}
        \label{fig:nba_c}
      \end{subfigure}

    \begin{subfigure}{\linewidth}
        \centering
        \includegraphics[width=\linewidth]{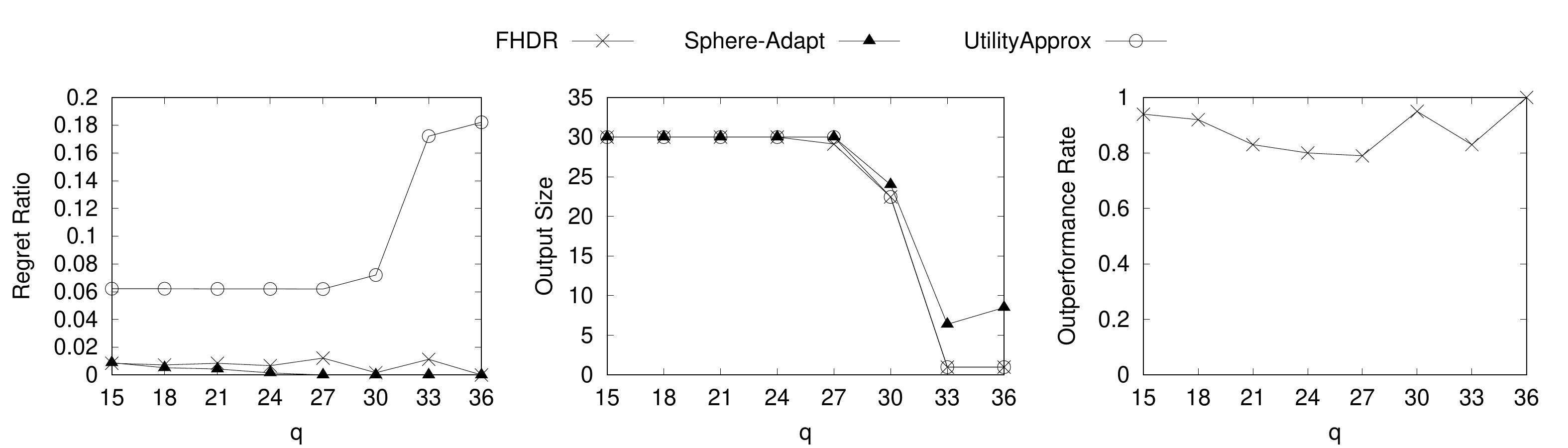}
        \vspace{-1.5em}
        \caption{P2: Vary $q$}
        \label{fig:nba_d}
      \end{subfigure}
    
    \begin{subfigure}{\linewidth}
        \centering
        \includegraphics[width=\linewidth]{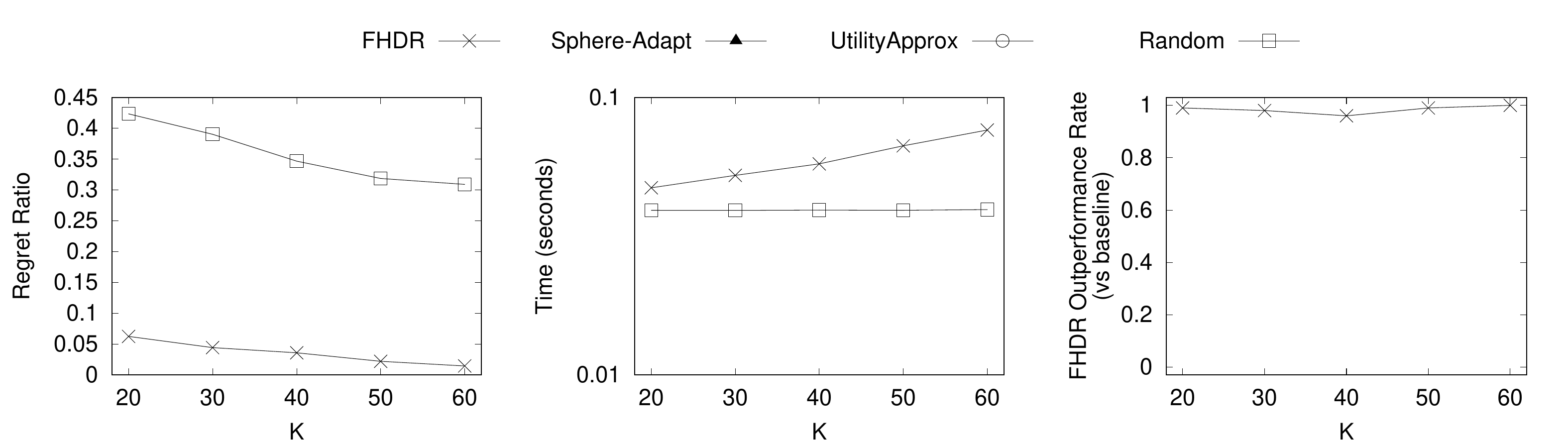}
        \vspace{-1.5em}
        \caption{$P2^{*}$: Vary $K$}
        \label{fig:p2star_nba}
    \end{subfigure}

    \caption{Results on \textit{NBA} dataset}
    \label{fig:nba}
\end{figure}

For the case where the user answers all questions (Problem~1 (\ref{p1})), as \(d_{\text{int}}\) increases from 2 to 5, FHDR requires 30–40 questions and approximately 0.04 s, while UtilityApprox requires 1,149 questions (Figure~\ref{fig:nba_a}).

Figures~\ref{fig:nba_b}--\ref{fig:p2star_nba} report results for Problem~2, including its zero-feedback special case. For the positive-feedback experiments, we set the default budget to $q=\lceil d/m\rceil$, sufficient to complete Phase 1, following~\cite{LA24} and the synthetic-data setting. For NBA, this gives $q=15$. When varying $q$, we consider values from $\lceil d/m\rceil$ to the maximum number of questions observed across repeated runs for identifying the user’s favorite tuple.

Figure~\ref{fig:nba_b} evaluates the effect of varying $d_{\text{int}}$. \textsc{FHDR} achieves regret ratios comparable to \textsc{Sphere-Adapt} and lower than \textsc{UtilityApprox}. Its execution time decreases slightly from $0.08$ to $0.05$ seconds and remains lower than both baselines. On NBA, the smaller tuple count makes the absolute runtime difference less pronounced.

As \(K\) increases, regret decreases for all methods (Figure~\ref{fig:nba_c}). AttributeSubset achieves an outperformance rate of approximately 0.9 and runs faster than Sphere-Adapt. FHDR maintains low regret with only $15$ questions and reaches zero regret with sufficient feedback (Figure~\ref{fig:nba_d}). \textsc{UtilityApprox}'s regret increase again coincides with the reduction to singleton output under the evaluation protocol described above.

\paragraph*{Zero-feedback special case ($P2^{*}$)}
As shown in Figure~\ref{fig:p2star_nba}, FHDR’s mean regret is lower than \textsc{Random} by an order of magnitude. The quality advantage is greater than on the synthetic dataset: FHDR achieves lower or equal regret in $96$--$100\%$ of runs, with execution time all below $0.1$s. Thus, the zero-feedback advantage is more pronounced on NBA, with a small additional computation cost.

Across the evaluated real datasets, FHDR combines low regret with efficient execution and reduced interaction cost, consistent with the synthetic-data results. These findings support its practical effectiveness on high-dimensional real data.

\subsection{User Study}
\label{sec:user_study}

\begin{figure}[ht]
  \centering
  \captionsetup[subfigure]{font=scriptsize,justification=centering}

  \begin{subfigure}[t]{0.48\columnwidth}
    \centering
    \includegraphics[width=\linewidth]{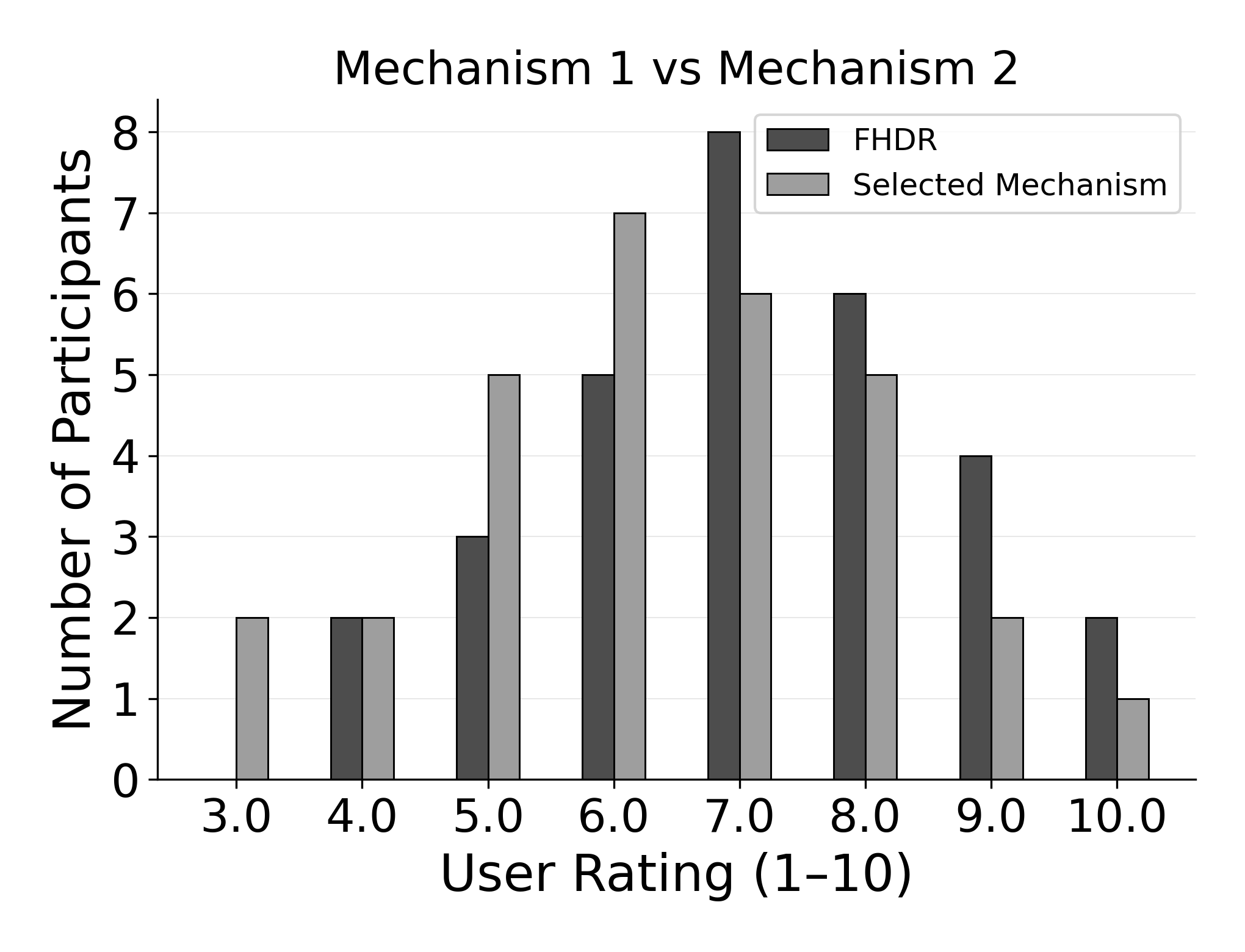}
    \caption{Mechanism 1 vs.~2}
    \label{fig:userstudy_2}
  \end{subfigure}
  \hfill
  \begin{subfigure}[t]{0.48\columnwidth}
    \centering
    \includegraphics[width=\linewidth]{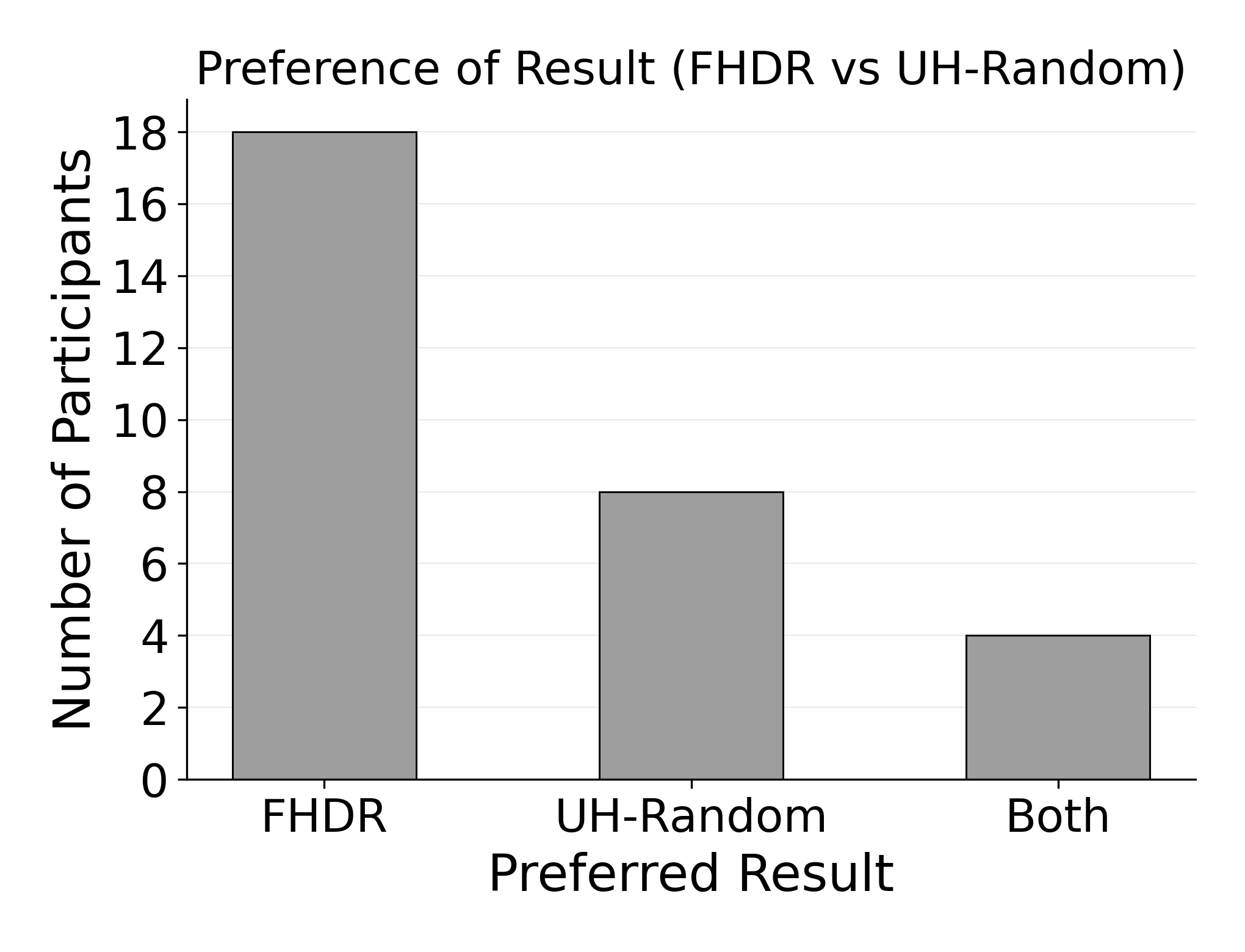}
    \caption{Result preference}
    \label{fig:userstudy_3}
  \end{subfigure}

  \caption{User study results}
  \label{fig:userstudy}
  \vspace{-0.8em}
\end{figure}

We conducted a user study\footnote{Study interface: \url{https://highdim-rm-demo-production.up.railway.app/}.} on the Car dataset to evaluate the practical usability of \textsc{FHDR}, because vehicle selection is intuitive and the dataset's $57$ attributes provide a representative high-dimensional setting. The study involved the same $30$ participants as the attribute-selection survey, and the results were averaged across participants.

We compare \textsc{FHDR} with \textsc{UH-Random}, evaluating their interaction mechanisms and returned recommendations. \textsc{UtilityApprox} is excluded because it requires over $1{,}000$ questions, making a direct user-study comparison impractical. Single-round methods such as \textsc{Sphere} are excluded from this interaction-experience comparison. To reduce bias, the interface does not reveal which method is ours.

Participants interact with both methods. During each tuple comparison, the interface displays only the attributes used in that question and their values. Participants are explicitly instructed to treat omitted attributes as identical, following the interaction model in Section~\ref{sec:def}. For \textsc{UH-Random}, they select between $2$ and $7$ relevant attributes and interact with the method on the projected dataset. They then compare the recommendations and interaction mechanisms. Mechanism~1, used by \textsc{FHDR}, relies on pairwise comparisons over a small number of displayed attributes. Mechanism~2, used by \textsc{UH-Random}, requires upfront attribute selection but generally fewer subsequent questions.

Figure~\ref{fig:userstudy} summarizes the results, with detailed statistics in \iftoggle{TechnicalReport}{Section~\ref{appen:user_study} of the \technicalreport}{Appendix~\ref{appen:user_study}}. Figure~\ref{fig:userstudy_2} shows that Mechanism~1 received a higher average rating than Mechanism~2, with scores of $7.10$ and $6.37$, respectively, while Mechanism~2 exhibited greater variation. These results support the usability advantage of avoiding upfront attribute selection. In Figure~\ref{fig:userstudy_3}, $18$ participants preferred FHDR’s output, $8$ preferred UH-Random’s, and $4$ rated them as comparable.

%% file: conclusions.tex
\section{Conclusion}
\label{sec:concl}

We presented FHDR for high-dimensional regret minimization under sparse linear preferences. FHDR discovers key attributes without upfront attribute selection and supports flexible interaction, including skipped questions, early stopping, and zero feedback. It identifies a favorite tuple with sufficient feedback and provides compact recommendations when feedback is limited. Experiments on synthetic and real datasets demonstrate low regret, efficient execution, and substantially reduced interaction cost, while the user study supports its practical usability. Future work could investigate error-tolerant interaction to handle inconsistent or unreliable responses.

\clearpage

%% file: appendix.tex
\clearpage
\appendices

\section{Summary of Notations}
\label{appen:summary}
We summarize the notations that are frequently used in Table \ref{tab:notations}.

\begin{table}[h]
\centering
\small
\renewcommand{\arraystretch}{1.15}
\setlength{\tabcolsep}{4pt}

\renewcommand{\arraystretch}{1.2}
\begin{tabular}{|>{\centering\arraybackslash}m{0.20\linewidth}|>{\centering\arraybackslash}m{0.72\linewidth}|}
\hline
\textbf{Notation} & \textbf{Meaning} \\
\hline\hline
$X$ & The set of $d$-dimensional points with $|X| = n$ \\
\hline
$X_\mathcal{D}$ & $X$ restricted to a selected subset $\mathcal{D}$ of dimensions\\
\hline
$f$ and $u$ & $f(p)=u\cdot p$ where $\sum_{i=1}^{d} u[i]=1$ \\
\hline
$f_\mathcal{D} (p)$ & Partial utility, $f_\mathcal{D} (p) = \sum_{i:D_i\in \mathcal{D}}u[i]p[i]$ \\
\hline
$\mathrm{rr}_X(S,u)$ & The regret ratio of $S$ over $X$ w.r.t.\ $u$ \\
\hline
$d$ & The number of dimensions of the original dataset \\
\hline
$d_\text{int}$ & The number of dimensions that the user is interested in \\
\hline
$d_\text{max}$ & The maximum number of dimensions that a user may be interested in \\
\hline
$d_\text{left}$ & The remaining upper bound on the number of key attributes in Phase 2 \\
\hline
$\mathcal{D}_\text{cand}$ & Candidate set: the set of dimensions which are not determined as \textit{non-key} \\
\hline
$D_i$ & The $i^{th}$ dimension from the original dataset, $i\in[1,d]$\\
\hline
$p^*$ & The point that a user like the most (having the maximum utility) \\
\hline
$S$ & The returned subset of points with $|S|=K$ \\
\hline
$K$ & The desired final output size for Problem 2 \\
\hline
$S_i$ & The returned regret minimizing set of \textsc{Sphere} in the $i^{th}$ iteration with $|S_i|=k$ \\
\hline
$k$ & The inner output size of \textsc{Sphere} in \textsc{AttributeSubset} \\
\hline
$m$ & The maximum number of attributes displayed per question \\
\hline
$w$ & The number of dimensions sampled in each iteration, in \textsc{AttributeSubset} \\
\hline
$N$ & The number of iterations executed in \textsc{AttributeSubset} \\
\hline
$s$ & The number of points displayed per question \\
\hline
$\varepsilon$ & The target regret ratio (in \textsc{UH-Random})\\
\hline
\end{tabular}
\caption{Summary of Frequently Used Notations}
\label{tab:notations}
\end{table}

\iftoggle{TechnicalReport}
  {\section{Remaining Proofs and Cap Analysis}}
  {\section{Remaining Proofs}}
\label{appen:proof}

\begin{proof}[Proof of Theorem~\ref{theorem_q}]
Let $Q(N,r)$ denote the worst-case number of remaining questions for standard generalized binary splitting when at most $r$ key attributes remain. We prove $Q(N,r)\leq B(N,r)$ by strong induction on $N+r$.

First, $B(N,r)$ is nondecreasing in $N$ for fixed $r$. Within each interval having fixed $\alpha$, this follows from $p=\lfloor l/2^\alpha\rfloor-r$; at an exponent transition, the bound increases by one. Moreover, in the third branch,
\[
\begin{aligned}
N-B(N,r)&=(2^\alpha-\alpha-1)r\\
&\quad +(2^\alpha-1)p+\theta\geq0,
\end{aligned}
\]
so $B(N,r)\leq N$.

If $N=0$ or $r=0$, no further question is required. If $1\leq N\leq2r-2$, individual testing uses at most $N=B(N,r)$ questions.

If $r=1$, then $B(N,1)=\lfloor\log_2N\rfloor+1=\alpha+1$. A positive group test followed by binary splitting uses at most $1+\alpha$ questions and identifies the only possible remaining key attribute. A negative test leaves $N-2^\alpha<2^\alpha$ candidates; induction bounds the total by $1+\alpha$.

If $r\geq2$ and $2r-1\leq N\leq3r-2$, then $\alpha=0$ and $B(N,r)=N$. The first question tests one attribute. Either response leaves at most $N-1$ candidates, so induction and $B(N',r')\leq N'$ bound the total by $N$.

It remains to consider $r\geq2$ and $N\geq3r-1$. Set $b=2^\alpha$. A negative response costs one question and leaves $(N-b,r)$. A positive response costs one group question and at most $\alpha$ splitting questions, identifies one key attribute, and leaves at most $N-1$ candidates with remaining upper bound $r-1$. Therefore,
\[
Q(N,r)\leq\max\{1+B(N-b,r),\;1+\alpha+B(N-1,r-1)\}.
\]
Substitution of the corrected decomposition gives
\[
1+B(N-b,r)=
\begin{cases}
B(N,r)-1,&p=0\ \text{and}\ \theta<b/2,\\
B(N,r),&\text{otherwise},
\end{cases}
\]
and
\[
1+\alpha+B(N-1,r-1)=
\begin{cases}
B(N,r)-1,&p=r-1,\\
B(N,r),&\text{otherwise}.
\end{cases}
\]
Both terms are at most $B(N,r)$, completing the induction.
\end{proof}

\paragraph{Capped FHDR procedure} Fix $m=7$. Define $C(0,r)=C(N,0)=0$, and $C(N,r)=N$ when $r\geq1$ and $1\leq N\leq2r-2$. Otherwise, let $g=\min\{2^\alpha,7\}$ and define
\[
\begin{aligned}
C(N,r)&=\max\{C_-(N,r),C_+(N,r)\},\\
C_-(N,r)&=1+C(N-g,r),\\
C_+(N,r)&=1+\lceil\log_2g\rceil\\
&\quad+\max_{1\leq j\leq g}C(N-j,r-1).
\end{aligned}
\]
A negative response removes all $g$ tested candidates. A positive response requires at most $\lceil\log_2g\rceil$ additional questions under balanced binary splitting and removes between one and $g$ candidates, including the identified key attribute. Thus, the two terms bound all possible responses. Every recursive state has fewer candidates, so strong induction on $N$ establishes $C$ as an upper bound for the capped procedure.

\paragraph{Finite verification of the FHDR bound} Evaluate the recurrence in increasing $N$ for $0\leq N\leq35$ and $0\leq r\leq5$. All dependencies refer to previously computed rows. The resulting $r=5$ column is
\[
C(N,5)=
\begin{cases}
N,&0\leq N\leq13,\\
7+\lfloor N/2\rfloor,&14\leq N\leq23,\\
13+\lfloor N/4\rfloor,&24\leq N\leq35.
\end{cases}
\]
These values equal $B(N,5)$ throughout the stated range. Consequently, $C(N,5)\leq17$ for $N\leq21$, and $C(N,5)\leq21$ for $N\leq35$. This establishes the stated FHDR bounds without applying the uncapped theorem to intermediate capped states.

\paragraph{Proof of Theorem \ref{theorem_batch}}

\begin{equation*}
\begin{aligned}
    \text{batch size}
    &= 2^\alpha = 2^{\lfloor\log_2 ((|\mathcal{D}_\text{cand}|- d_\text{left}+1)/d_\text{left})\rfloor} \\
    &= 2^{\lfloor\log_2 ((|\mathcal{D}_\text{cand}|-d_\text{max}+1)/d_\text{max})\rfloor}\\
    &\leq (|\mathcal{D}_\text{cand}|-d_\text{max}+1)/d_\text{max}\\
    &\leq |\mathcal{D}_\text{cand}|/d_\text{max} \leq d_\text{int}m/d_\text{max}\\
    &\leq m
\end{aligned}
\end{equation*}

\begin{proof}[Proof of Theorem~\ref{theorem_att}]
Write $p_{\mathcal{D}_w}$ for the projection of $p$, and let $\bar u$ contain the coefficients of $u$ on $\mathcal{D}_w$. Coverage implies that every omitted coefficient is zero. Thus, $\bar u\geq0$, $\lVert\bar u\rVert_1=1$, and $u\cdot p=\bar u\cdot p_{\mathcal{D}_w}$ for every $p\in X$. Consequently,
\[
\begin{aligned}
\mathrm{rr}_X(S_i,u)
&=1-\frac{\max_{p\in S_i}\bar u\cdot p_{\mathcal{D}_w}}
{\max_{p\in X}\bar u\cdot p_{\mathcal{D}_w}}\\
&=\mathrm{rr}_{X_{\mathcal{D}_w}}\bigl((S_i)_{\mathcal{D}_w},\bar u\bigr)\\
&\leq\varepsilon.
\end{aligned}
\]
The last inequality follows from the assumed projected regret bound.
\end{proof}

\paragraph{Trimming retention}
Fix $u$, and condition on the history before postprocessing. For an oversized union of size $M$, the preceding union contains at most $K-1$ tuples and the last run adds at most $k$, so $M\leq K+k-1$. Uniform trimming retains a specified maximizer of $u\cdot p$ in $U$ with probability $K/M\geq\eta$, where $\eta=K/(K+k-1)$; filling or retaining an undersized or exact-size union preserves every tuple. Thus, writing $G=\{\mathrm{rr}_X(U,u)\leq\varepsilon\}$ and $\mathcal{F}$ for this history,
\[
\begin{aligned}
\Pr\bigl(\mathrm{rr}_X(S,u)\leq\varepsilon\bigr)
&\geq\mathbb{E}\!\left[\mathbf{1}_G\Pr(z\in S\mid\mathcal{F})\right]\\
&\geq\eta\Pr(G),
\end{aligned}
\]
where $z\in\arg\max_{p\in U}u\cdot p$ is chosen from the history using a fixed rule. Retaining $z$ preserves the union's regret bound.

\begin{proof}[Proof of Lemma~\ref{lemma_p}]
There are $\binom Dw$ equally likely samples. A sample containing $\mathcal{H}$ is completed by choosing $w-h$ attributes from the remaining $D-h$ attributes. Hence,
\[
\begin{aligned}
P_\text{cover}&=\frac{\binom{D-h}{w-h}}{\binom Dw}\\
&=\frac{\binom wh}{\binom Dh}\\
&=\prod_{j=0}^{h-1}\frac{w-j}{D-j}\\
&\geq\left(\frac{w-h+1}{D}\right)^h.
\end{aligned}
\]
The last inequality follows from $w-j\geq w-h+1$ and $D-j\leq D$ for every $0\leq j<h$.
\end{proof}

\paragraph{Fixed-round coverage}
Each sample fails to cover the key attributes with probability $1-P_\text{cover}$. Independence therefore gives $\Pr(C_t^c)=(1-P_\text{cover})^t$, proving the coverage formula. Solving $(1-P_\text{cover})^t\leq1-\gamma$ gives $t\geq\ln(1-\gamma)/\ln(1-P_\text{cover})$. Taking the ceiling yields the minimum integer $t_\gamma$.

\iftoggle{TechnicalReport}
  {\section{Capability Qualifications and Reproduction Results}}
  {\section{Capability Qualifications}}
\label{appen:capabilities}

\paragraph{Comparison criteria}
In Table~\ref{tab:capabilities}, practical high-dimensional support concerns practical computation and interaction costs. A half-check denotes substantial cost or a required implementation repair; a cross denotes execution failure in the tested setting. Execution assessments concern the tested \textsc{HD-PI} and \textsc{Verify-Point} implementations; \textsc{RH} and \textsc{Verify-Space} remain untested.

\paragraph{Exact-favorite guarantee}
Exact-favorite guarantees assume sufficient feedback: \textsc{HD-PI}/\textsc{RH} use target rank one for favorite retrieval~\cite{WWX21}, \textsc{Verify-Point}/\textsc{Verify-Space} provide probabilistic guarantees under their error model~\cite{CW23}, and FHDR assumes sparse preferences.

\paragraph{Implicit key-attribute discovery}
Key-attribute discovery excludes learning weights over an upfront user-selected attribute set.

\paragraph{Output size control}
Output size control means a user-specified output budget independent of the original dimensionality. Published set-output extensions are included; fixed singleton output is excluded. \textsc{Sphere} assumes $k\geq d$~\cite{XW18}. The original \textsc{UtilityApprox} algorithm returns a singleton and does not specify arbitrary-size final output~\cite{NL12}. The \textsc{HD-PI}/\textsc{RH} entries include the published \textsc{AllTopK} and \textsc{SomeTopK} extensions~\cite{WWX21}; the \textsc{Verify-Point} size-control entry includes the published set-output extensions~\cite{CW23}. FHDR's output size control applies before interactive Phase~3 through \textsc{AttributeSubset}; during Phase~3, it returns one recommendation. \textsc{Breakpoint} returns an indistinguishability candidate set whose cardinality is not prescribed by the user~\cite{LA25}.

\paragraph{Output at any interaction budget}
This criterion requires a specified recommendation procedure at every interaction budget, including zero feedback. Recommendation quality depends on the available information. Early stopping alone does not establish zero-feedback support.

\paragraph{Existing execution and interaction limitations}
\textsc{Sphere} exceeds one hour on the synthetic dataset with $d=100$ and $n=10^5$ and requires output size $k\geq d$. \textsc{UtilityApprox} requires approximately $1{,}100$ and $1{,}149$ questions on this synthetic dataset and NBA, respectively, making interaction impractical. Reproduced \textsc{UH-Random}/\textsc{UH-Simplex} and \textsc{HD-PI} encounter numerical failures during convex-hull computation on the targeted high-dimensional inputs.

\paragraph{Additional reproduction setting}
We tested the released \textsc{Breakpoint} and \textsc{Verify-Point} implementations at commits \texttt{adf0740} and \texttt{22016b7}, respectively, on an Apple M2 Pro with $32$~GiB RAM, seed $42$, and a one-hour deadline per main attempt. Inputs retained all original dimensions. Times include loading and native preprocessing. These are individual execution diagnostics; concurrent independent runs preclude interpreting the times as comparative speedups.

\paragraph{Breakpoint}
With $s=2$, indistinguishability threshold $0.05$, and $q=5d$, \textsc{Breakpoint} completes on NBA ($4{,}790\times104$) in $7.27$~s using $520$ rounds and returns $532$ tuples. With $30$ or zero rounds, it returns $3{,}813$ tuples. On the $100{,}000\times100$ synthetic input, a sanitizer confirms overflow in the published 32-bit slope counter: a tested count is $2{,}493{,}946{,}605>2^{31}-1$. A separate copy with 64-bit counts and corresponding sentinels, retaining the algorithmic procedures, completes in $747.04$~s using $500$ rounds and returns $18{,}794$ tuples. The original code completes in $1{,}039.12$~s but contains the overflow and returns all $100{,}000$ tuples. On the $10{,}000\times100$ input, both versions return all $10{,}000$ tuples. The repair, runtime, interaction burden, and large outputs motivate the half-check. Its additive indistinguishability-error diagnostic is distinct from regret ratio.

\paragraph{Verify-Point}
Using the supplied driver's error rate $0.05$ and checking count $3$, \textsc{Verify-Point} returns no recommendation on either principal input. NBA logs Qhull allocation error \texttt{QH6235} after generating approximately $7.7$ million facets. The NBA and $100{,}000\times100$ synthetic attempts are stopped during hull construction after $697.90$~s and $794.68$~s, with measured memory footprints of $17.7$ and $15.4$~GiB, respectively. These are termination times, not completed runs or one-hour timeouts. A supplied $100\times100$ input triggers feasible-point parsing error \texttt{QH6072} followed by a segmentation fault with both error-free and $5\%$ erroneous responses. Library paths were adapted and timing drivers added; the tested algorithm and geometry routines were unchanged. These results characterize the released implementations and tested settings.

\section{Additional Experimental Results on Real Datasets}
\label{appen:real}

\begin{figure*}[tbp]
  \centering
  \begin{subfigure}{0.48\linewidth}
    \centering
    \includegraphics[width=0.67\linewidth]{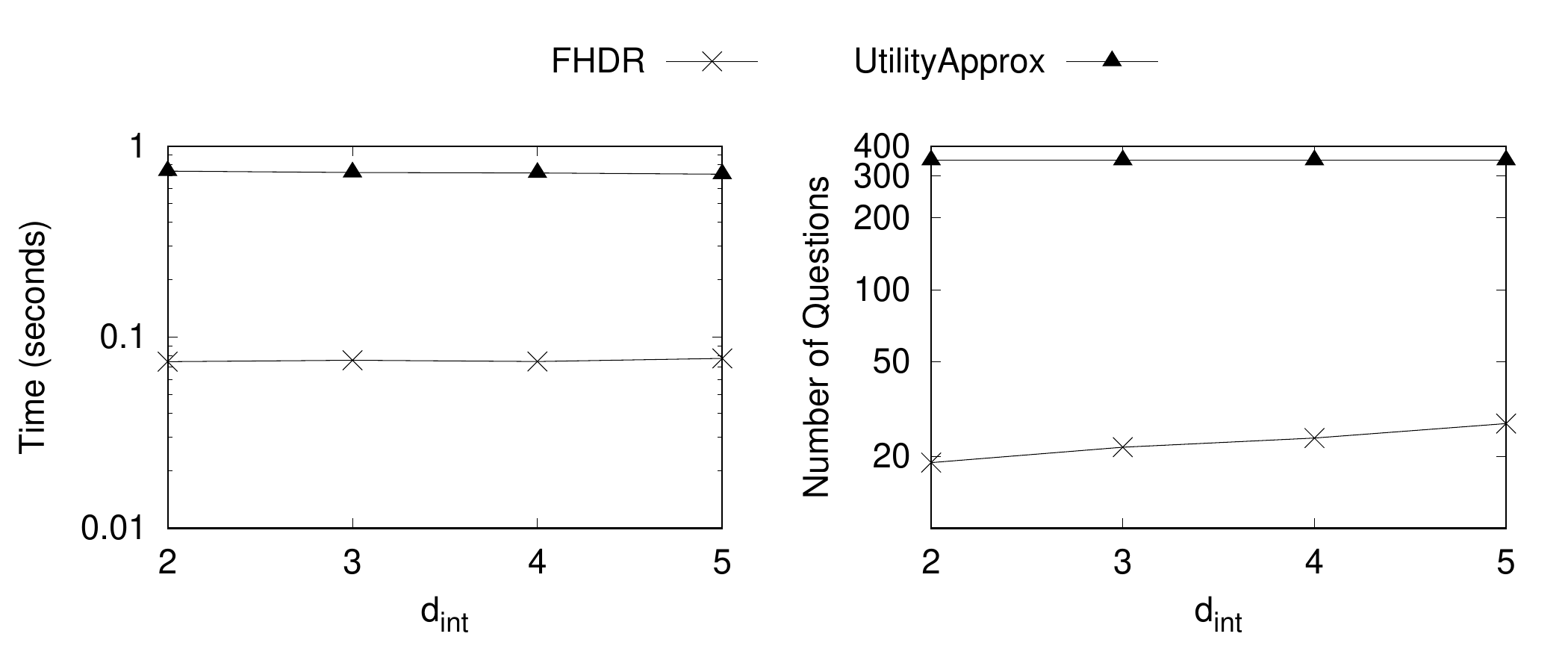}
    \caption{P1: Vary $d_{\text{int}}$}
    \label{fig:house_a}
  \end{subfigure}
  \hfill
  \begin{subfigure}{0.48\linewidth}
    \centering
    \includegraphics[width=\linewidth]{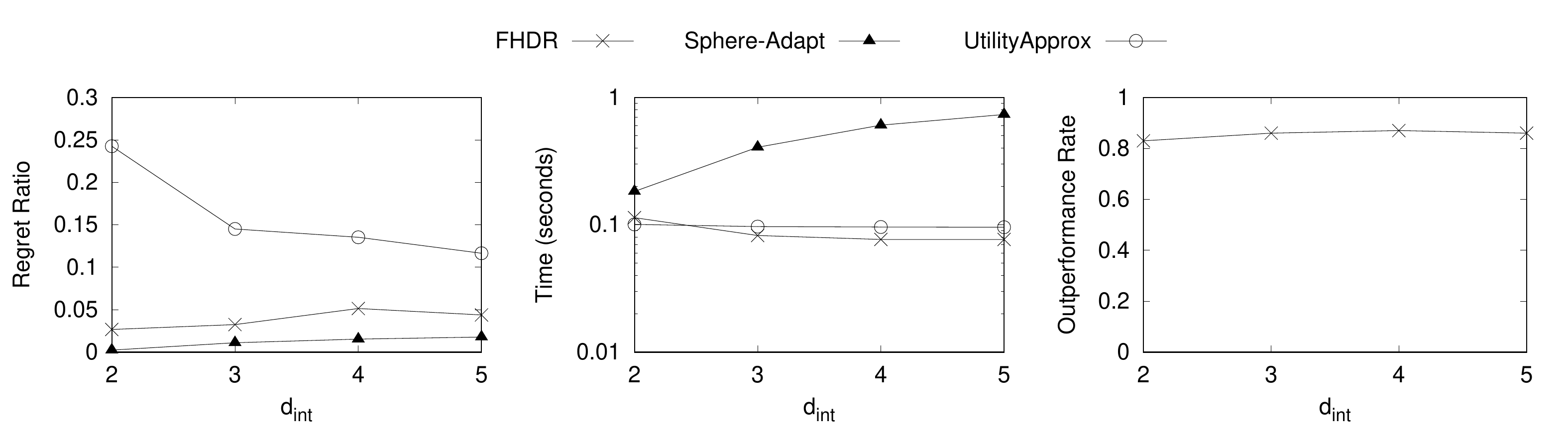}
    \caption{P2: Vary $d_{\text{int}}$}
    \label{fig:house_b}
  \end{subfigure}

  \vspace{0.6em}
  \begin{subfigure}{0.48\linewidth}
    \centering
    \includegraphics[width=\linewidth]{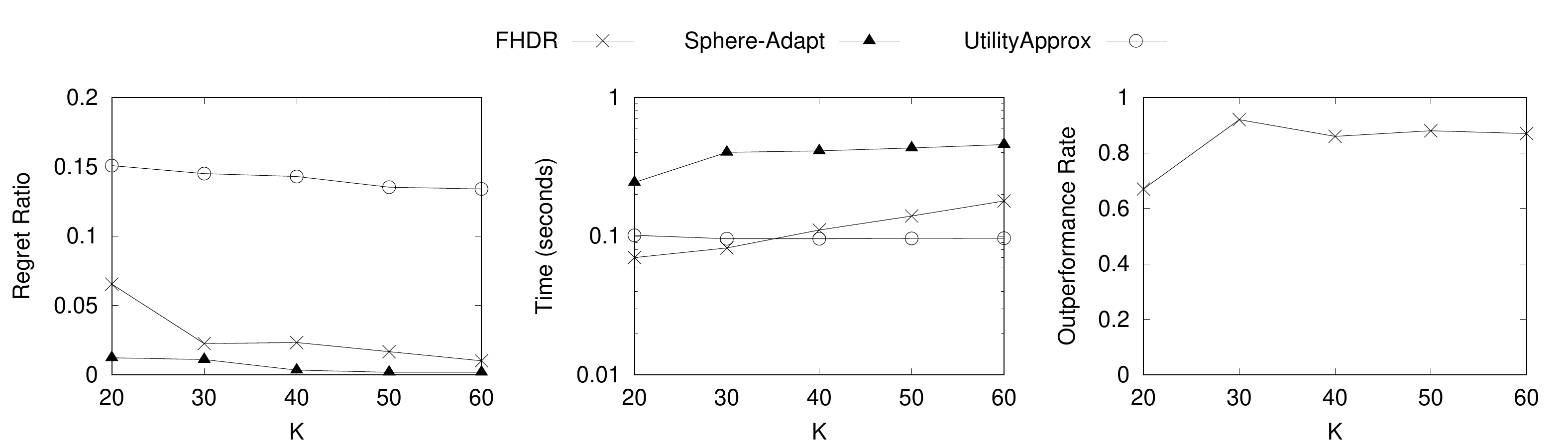}
    \caption{P2: Vary $K$}
    \label{fig:house_c}
  \end{subfigure}
  \hfill
  \begin{subfigure}{0.48\linewidth}
    \centering
    \includegraphics[width=\linewidth]{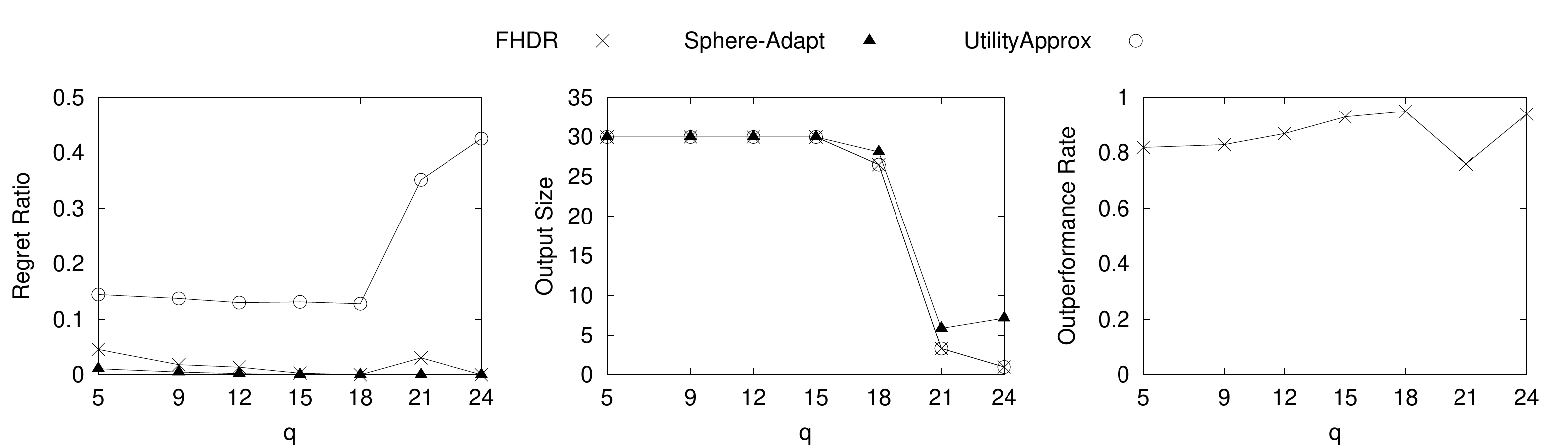}
    \caption{P2: Vary $q$}
    \label{fig:house_d}
  \end{subfigure}

  \caption{Results on \textit{House} dataset}
  \label{fig:house}
\end{figure*}

\begin{figure*}[tbp]
  \centering
  \begin{subfigure}{0.48\linewidth}
    \centering
    \includegraphics[width=0.67\linewidth]{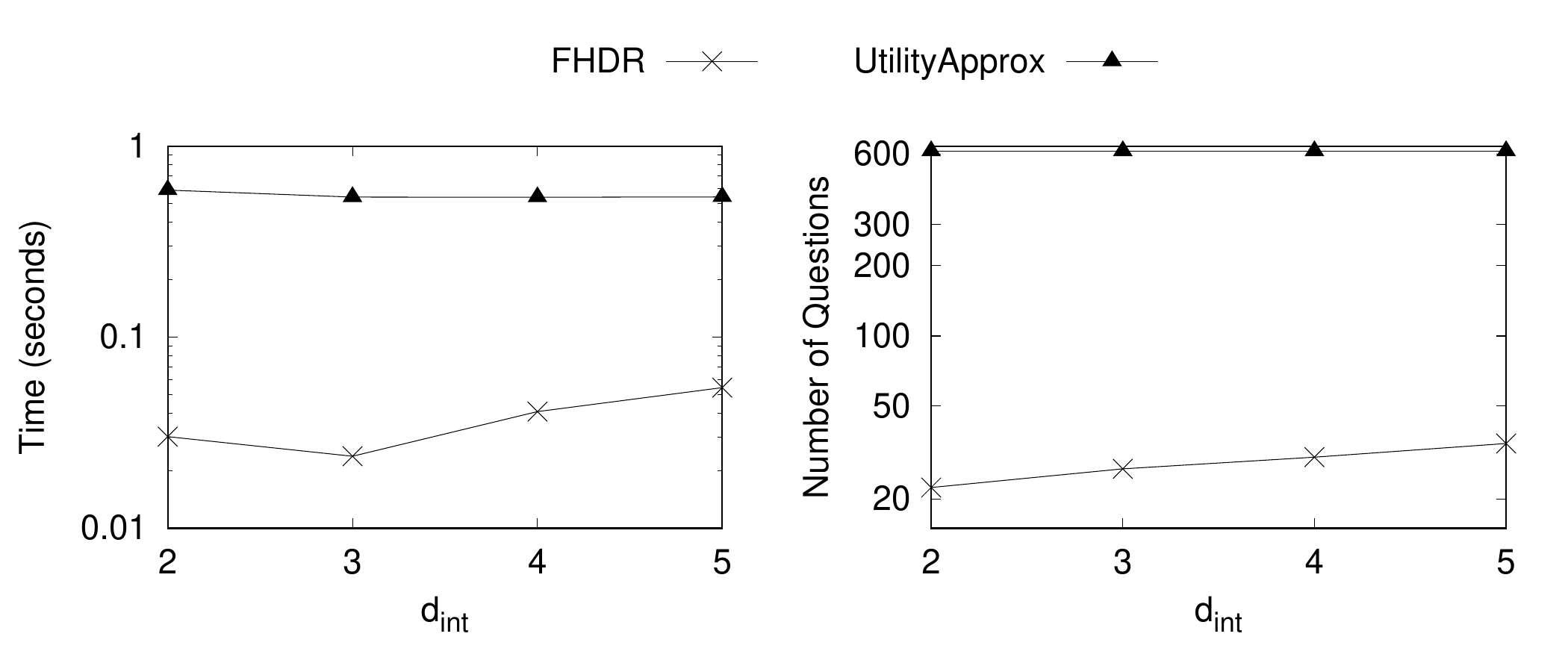}
    \caption{P1: Vary $d_{\text{int}}$}
    \label{fig:car_a}
  \end{subfigure}
  \hfill
  \begin{subfigure}{0.48\linewidth}
    \centering
    \includegraphics[width=\linewidth]{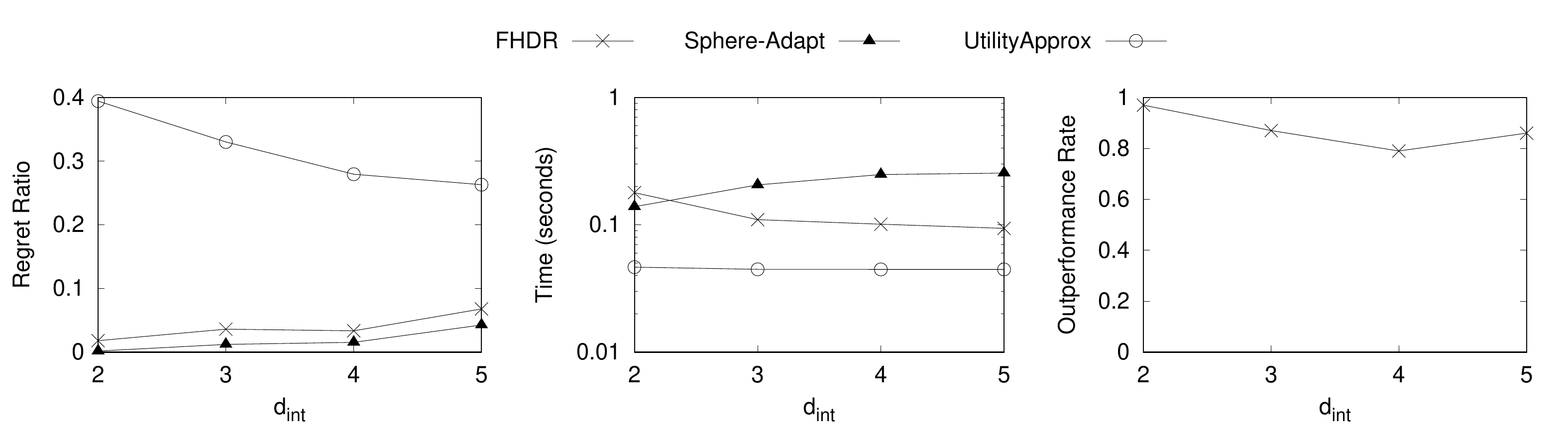}
    \caption{P2: Vary $d_{\text{int}}$}
    \label{fig:car_b}
  \end{subfigure}

  \vspace{0.6em}
  \begin{subfigure}{0.48\linewidth}
    \centering
    \includegraphics[width=\linewidth]{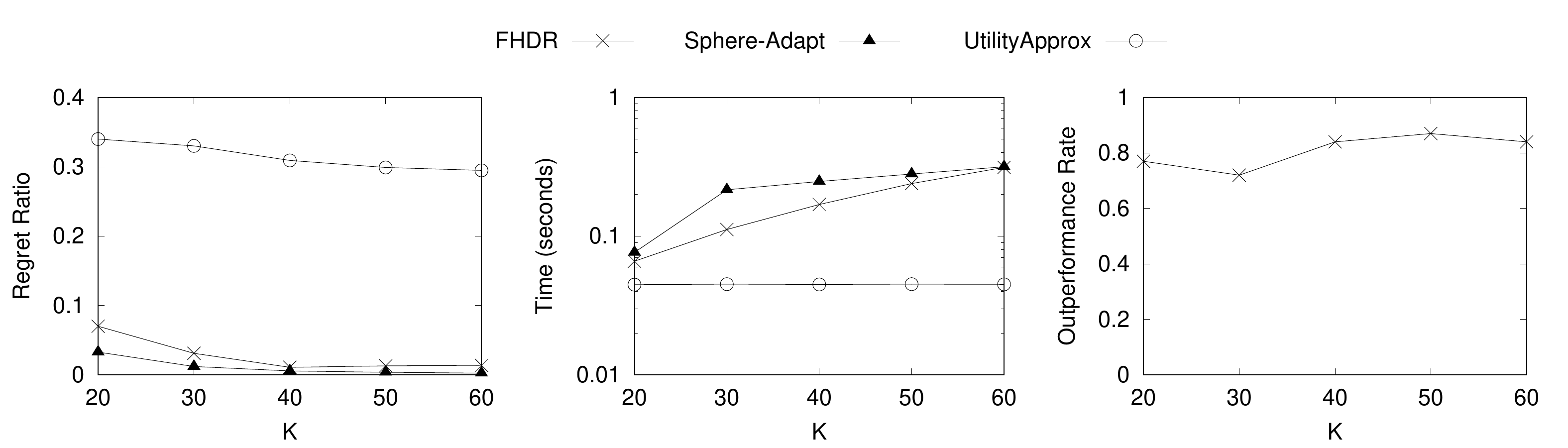}
    \caption{P2: Vary $K$}
    \label{fig:car_c}
  \end{subfigure}
  \hfill
  \begin{subfigure}{0.48\linewidth}
    \centering
    \includegraphics[width=\linewidth]{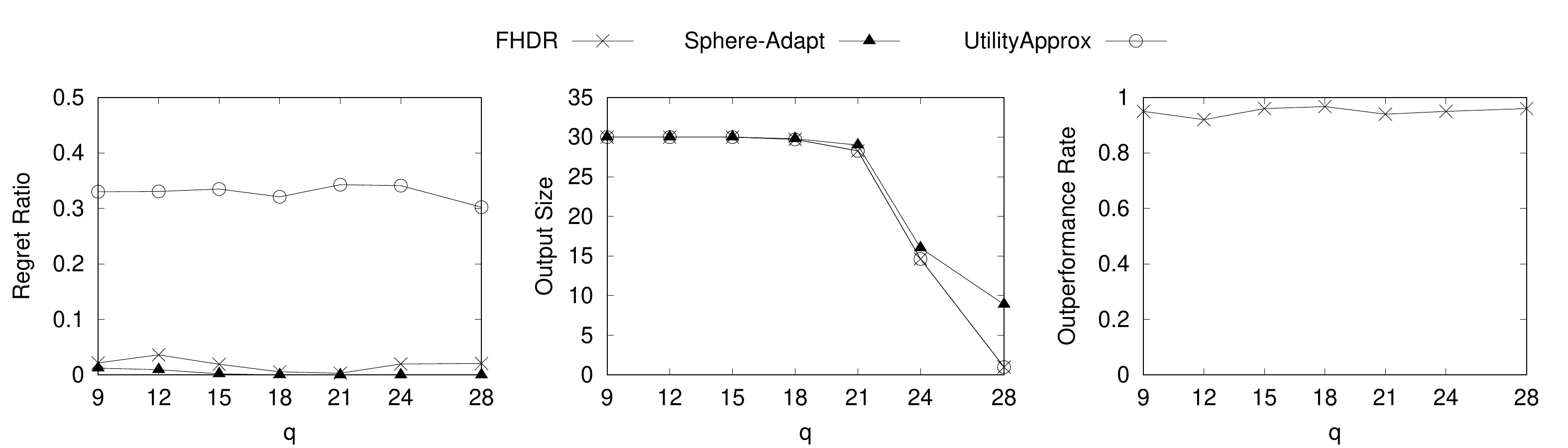}
    \caption{P2: Vary $q$}
    \label{fig:car_d}
  \end{subfigure}

  \caption{Results on \textit{Car} dataset}
  \label{fig:car}
\end{figure*}

\begin{figure*}[tbp]
  \centering
  \begin{subfigure}{0.48\linewidth}
    \centering
    \includegraphics[width=0.67\linewidth]{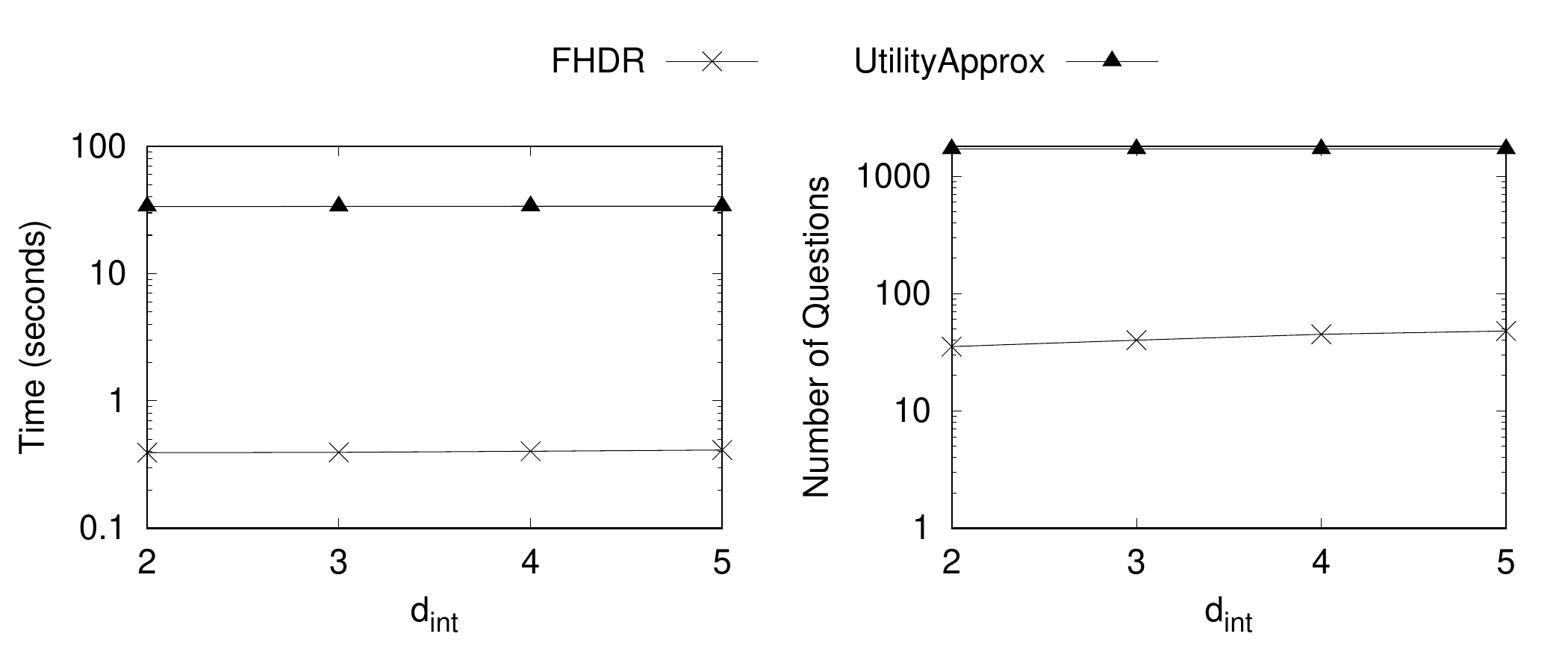}
    \caption{P1: Vary $d_{\text{int}}$}
    \label{fig:energy_a}
  \end{subfigure}
  \hfill
  \begin{subfigure}{0.48\linewidth}
    \centering
    \includegraphics[width=\linewidth]{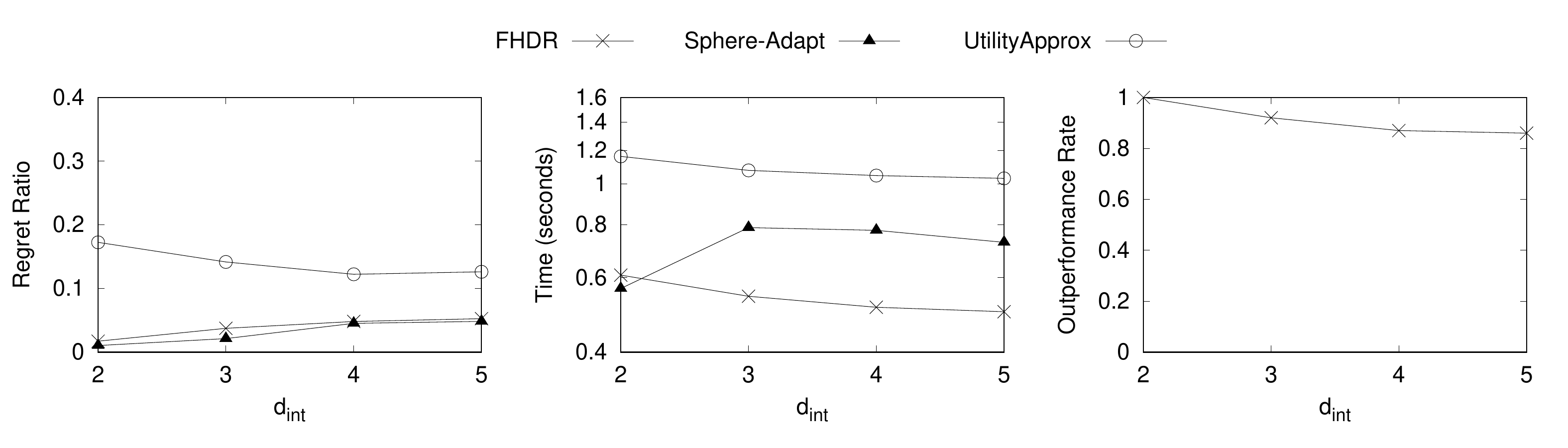}
    \caption{P2: Vary $d_{\text{int}}$}
    \label{fig:energy_b}
  \end{subfigure}

  \vspace{0.6em}
  \begin{subfigure}{0.48\linewidth}
    \centering
    \includegraphics[width=\linewidth]{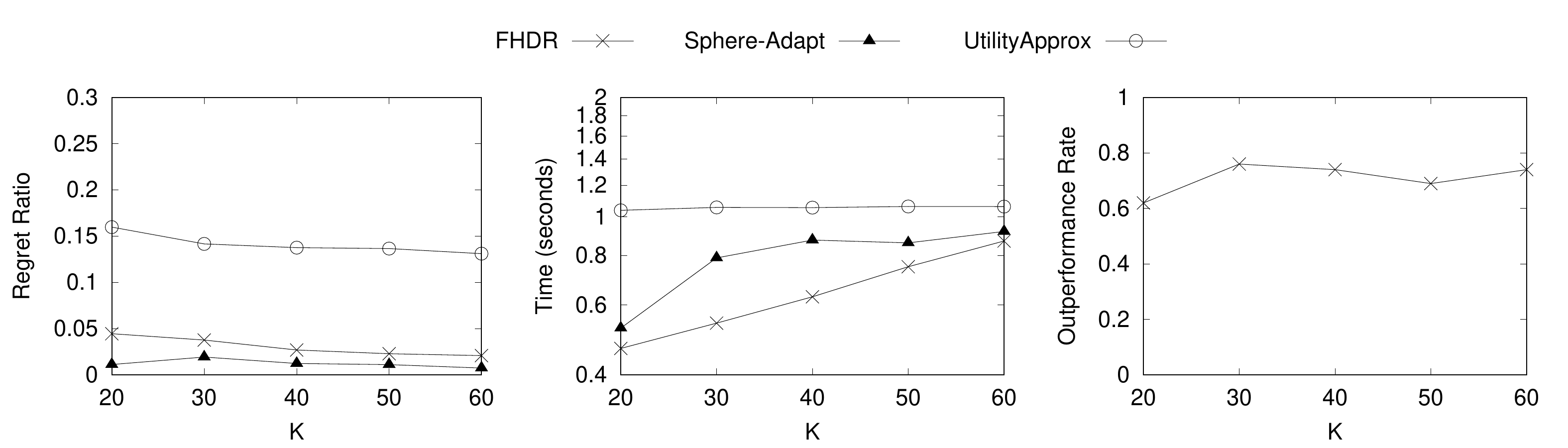}
    \caption{P2: Vary $K$}
    \label{fig:energy_c}
  \end{subfigure}
  \hfill
  \begin{subfigure}{0.48\linewidth}
    \centering
    \includegraphics[width=\linewidth]{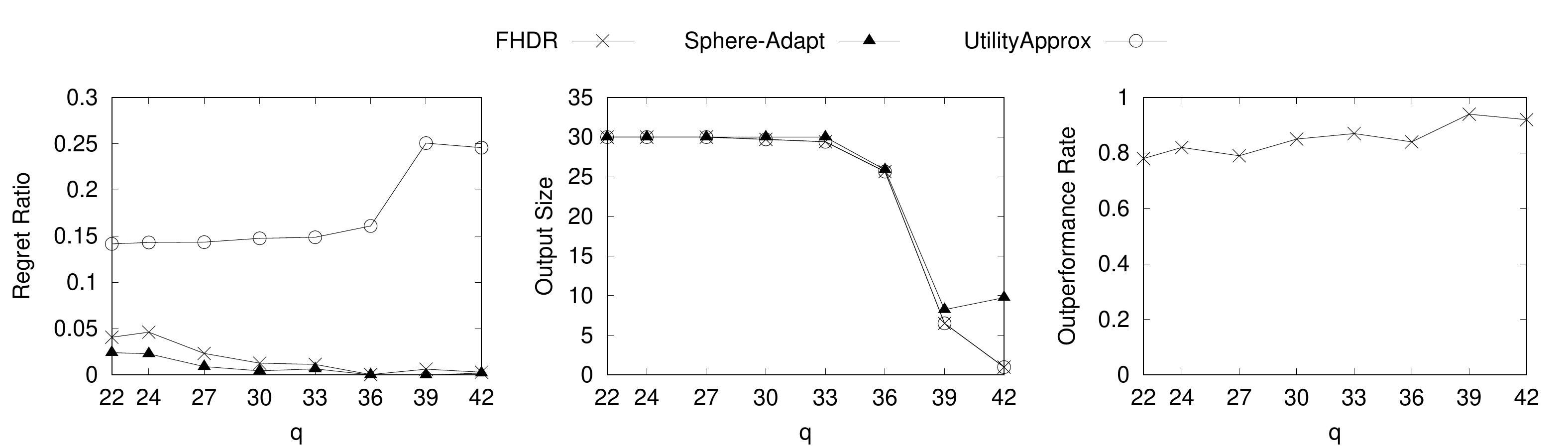}
    \caption{P2: Vary $q$}
    \label{fig:energy_d}
  \end{subfigure}

  \caption{Results on \textit{Energy} dataset}
  \label{fig:energy}
\end{figure*}

\begin{figure*}[tbp]
  \centering
  \begin{subfigure}{0.48\linewidth}
    \centering
    \includegraphics[width=0.67\linewidth]{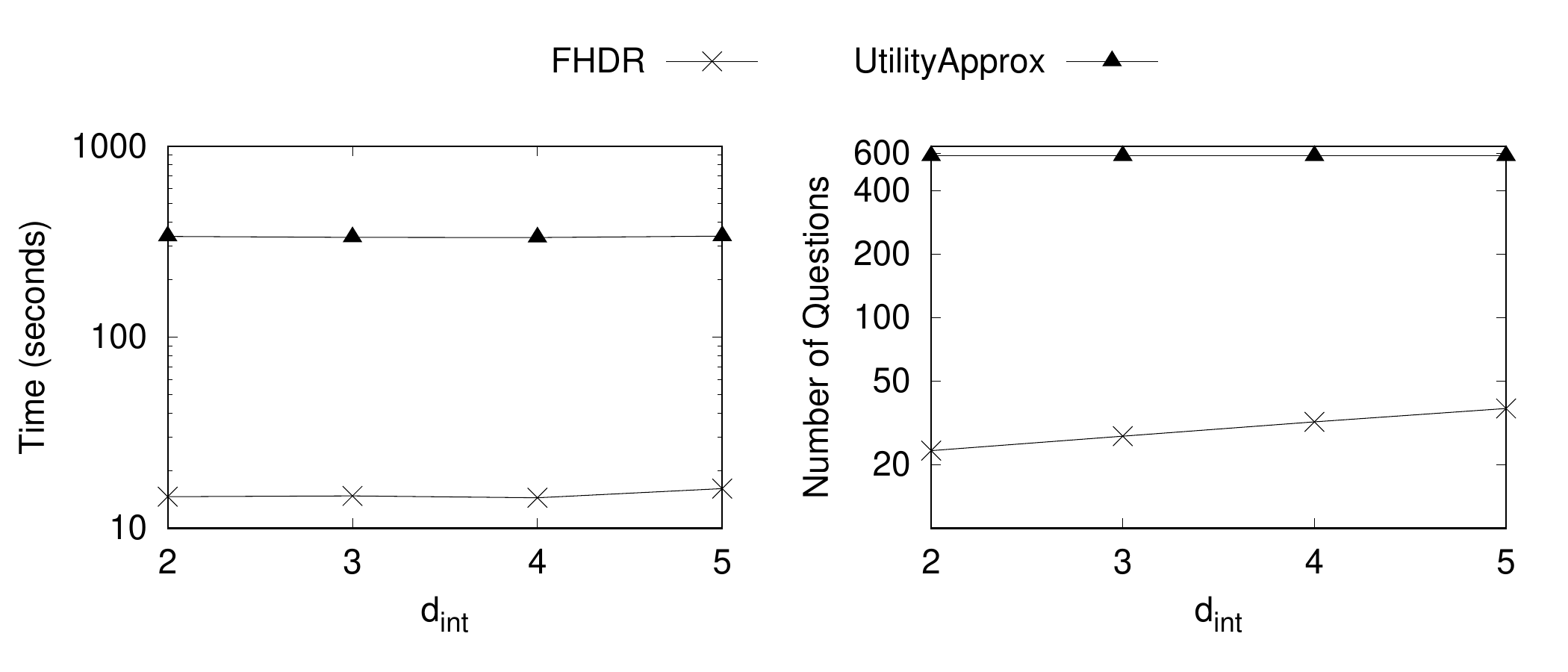}
    \caption{P1: Vary $d_{\text{int}}$}
    \label{fig:activity_a}
  \end{subfigure}
  \hfill
  \begin{subfigure}{0.48\linewidth}
    \centering
    \includegraphics[width=\linewidth]{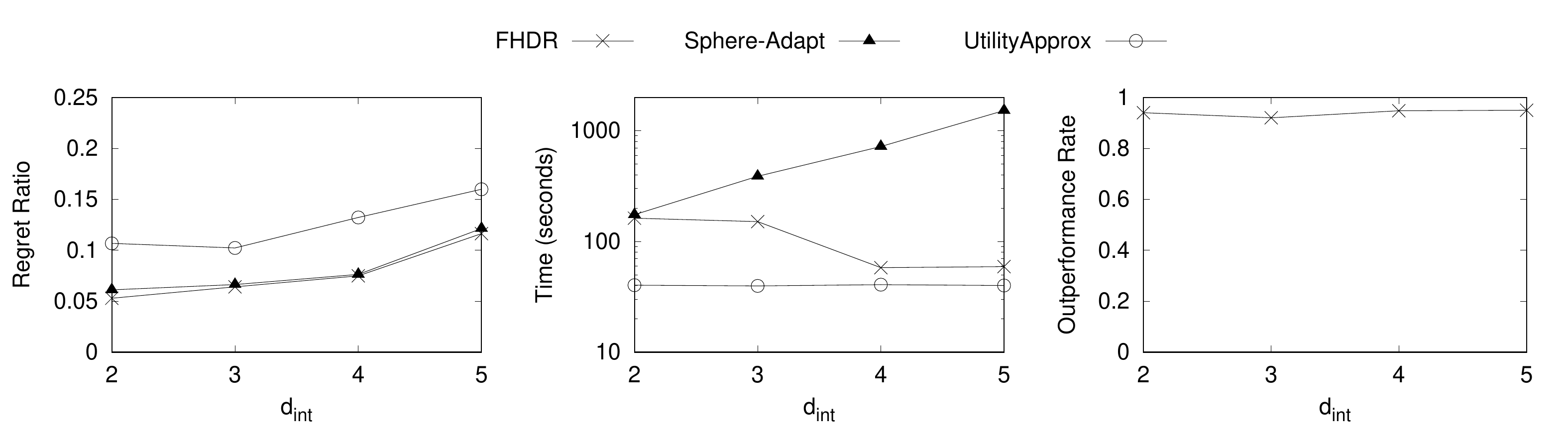}
    \caption{P2: Vary $d_{\text{int}}$}
    \label{fig:activity_b}
  \end{subfigure}

  \vspace{0.6em}
  \begin{subfigure}{0.48\linewidth}
    \centering
    \includegraphics[width=\linewidth]{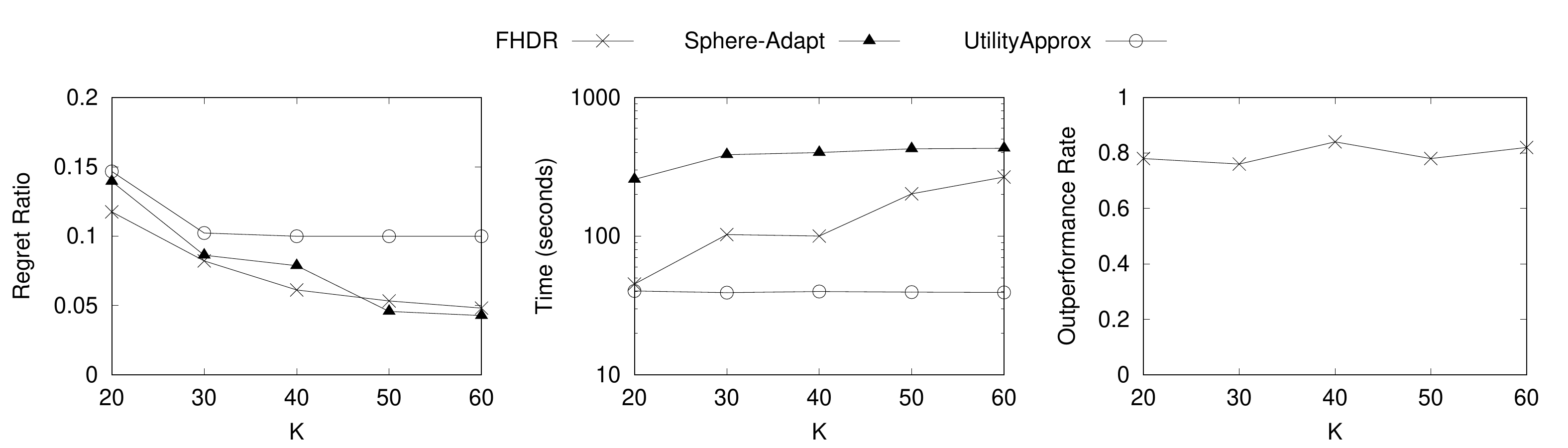}
    \caption{P2: Vary $K$}
    \label{fig:activity_c}
  \end{subfigure}
  \hfill
  \begin{subfigure}{0.48\linewidth}
    \centering
    \includegraphics[width=\linewidth]{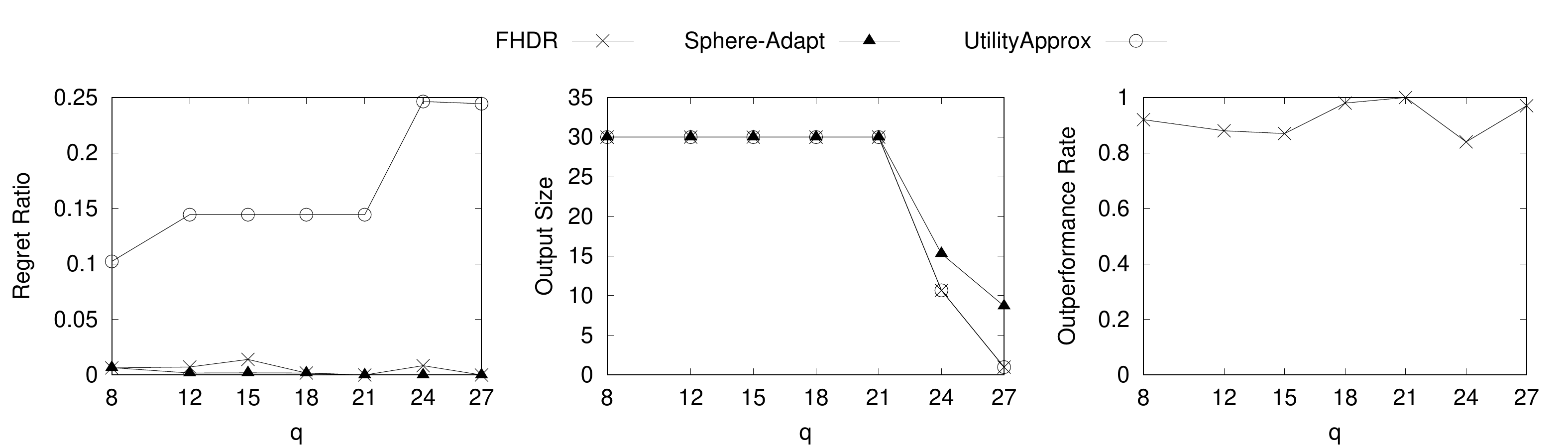}
    \caption{P2: Vary $q$}
    \label{fig:activity_d}
  \end{subfigure}

  \caption{Results on \textit{Activity} dataset}
  \label{fig:activity}
\end{figure*}

We present here the additional experimental results of four real datasets, namely House, Car, Energy and Activity. The default values of $q$ are $5$, $9$, $22$, $8$ respectively. The results can be found in Figures~\ref{fig:house}, \ref{fig:car}, \ref{fig:energy} and \ref{fig:activity}.

\section{Supplementary Material for Experimental Setting}
\label{appen:setting}

\usetikzlibrary{positioning}

\begin{figure*}[tbp]
  \centering
  \begin{tikzpicture}
    \node[inner sep=0] (img) {\includegraphics[width=\linewidth]{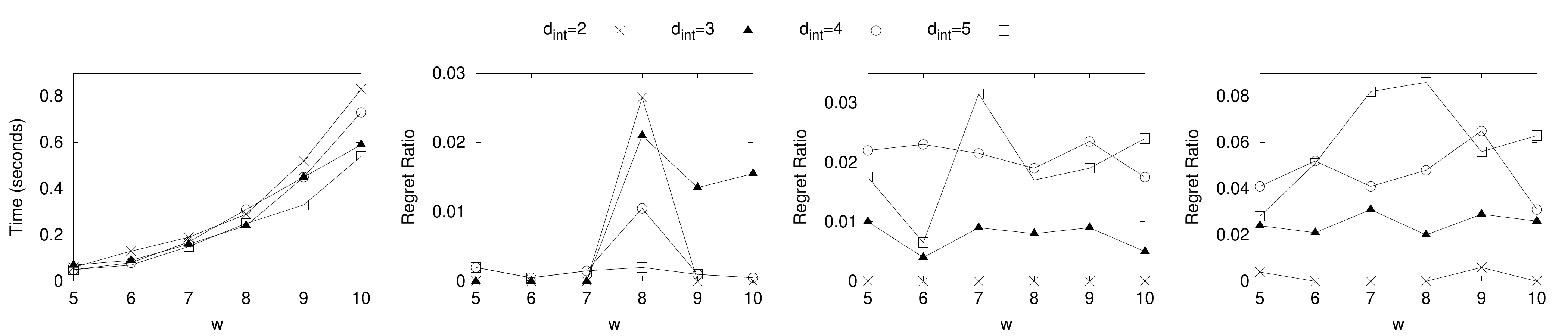}};

    \path (img.south west) -- (img.south east)
      coordinate[pos=.135] (c1)
      coordinate[pos=.395] (c2)
      coordinate[pos=.643] (c3)
      coordinate[pos=.89] (c4);

    \node[below=2mm of c1, anchor=north, xshift=0pt]  {(Energy)};
    \node[below=2mm of c2, anchor=north, xshift=0pt]  {(Car)};
    \node[below=2mm of c3, anchor=north, xshift=0pt]  {(NBA)};
    \node[below=2mm of c4, anchor=north, xshift=0pt]  {(Energy)};
  \end{tikzpicture}

  \caption{Vary $w$}
  \label{fig:w}
\end{figure*}

\begin{figure}[tbp]
  \centering

  \begin{minipage}[t]{0.54\linewidth}
    \centering
    \renewcommand{\arraystretch}{1.2}
    \vspace{0pt}
    \begin{tabular}{|c|c|c|}
      \hline
      Dataset & $d$ & $|X|$ \\
      \hline\hline
      House  & 33  & 15{,}171 \\
      \hline
      Car    & 57  & 5{,}500 \\
      \hline
      NBA    & 104 & 4{,}790 \\
      \hline
      Energy & 149 & 36{,}043 \\
      \hline
      Activity & 54 & 2{,}872{,}533 \\
      \hline
    \end{tabular}
    \captionof{table}{Real Datasets}
    \label{tab:real-datasets}
  \end{minipage}
  \hfill
  \begin{minipage}[t]{0.42\linewidth}
    \centering
    \vspace{0pt}
    \includegraphics[width=\linewidth]{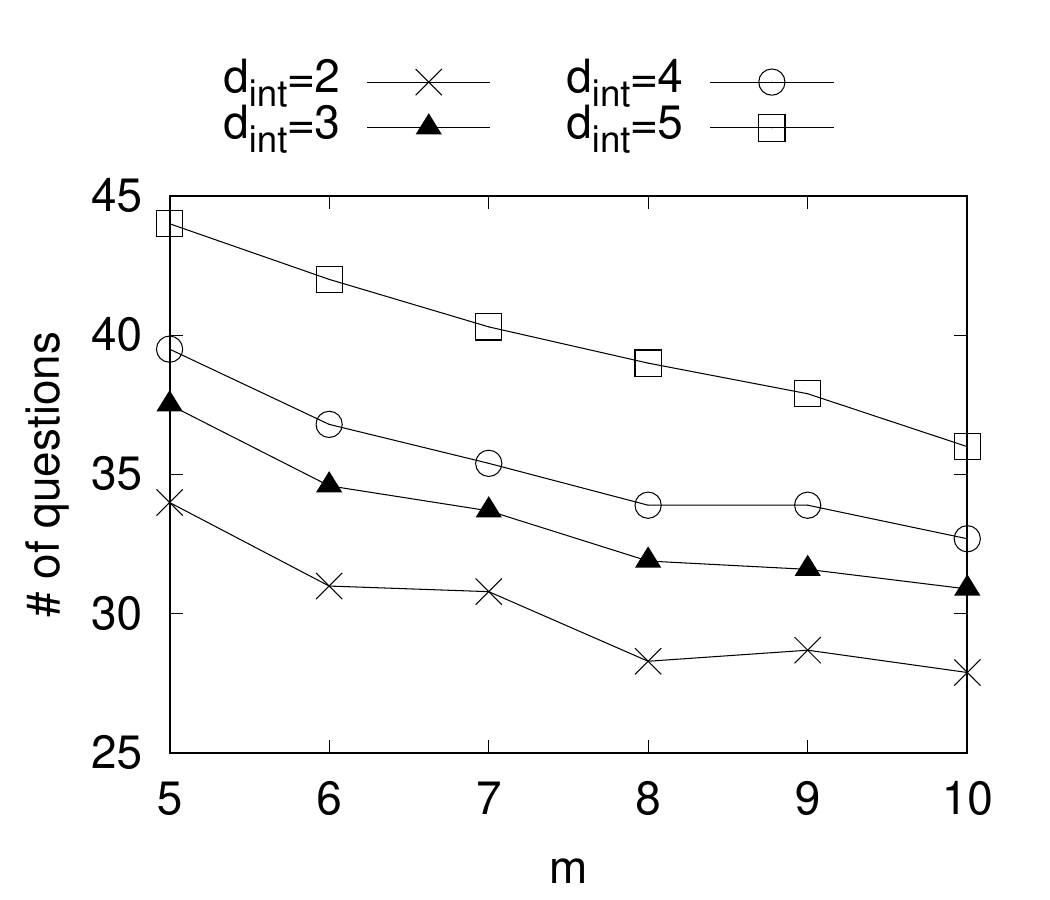}
    \captionof{figure}{Vary $m$}
    \label{fig:m}
  \end{minipage}

\end{figure}

We evaluate internal parameter choices in terms of regret ratio and execution time.

To assess $m$, we measure the number of questions required to identify the single tuple with the highest utility. As observed in Figure \ref{fig:m}, the number of questions required to identify the optimal tuple decreases as $m$ increases, since each question yields more information. However, a larger $m$ also makes each question more cognitively demanding for the user. We must balance these factors, as both the number of questions and the number of attributes per question contribute to overall cognitive load. Cognitive science literature indicates that humans can only comfortably process a limited number of items at once (e.g., $7$ as suggested by \cite{MG56}), and recent working memory studies find capacity constraints on the order of $4\sim 7$\cite{Morra2024}. In parallel, marketing research guidelines for conjoint analysis recommend using roughly $5\sim7$ attributes at most in each task to avoid overwhelming respondents \cite{McCullough2007}, also supported by some consumer research \cite{OA14, GS78}. Therefore, we use $m$ at 7 in our approach, as this upper limit balances query efficiency in our algorithm with user cognitive limits, aligning with the above guidelines.

For $w$, we restrict the number of questions so that the algorithm can only finish Phase 1, to evaluate performance in \textsc{AttributeSubset}. We plot the execution time on the Energy dataset, and the regret ratio on three datasets (from left to right): Car, NBA, Energy. Results are presented in Figure \ref{fig:w}. As shown in the first plot, the execution time increases sharply once $w$ exceeds $6$. Regarding the regret ratio, a larger $w$ raises the likelihood of including the \textit{key} attributes within each sampled subset, while simultaneously reducing the accuracy upon successfully including all \textit{key} attributes. These two effects counterbalance each other, leading to a more intricate, non-monotonic curve rather than a steady trajectory. In particular, there is a local valley at $w=6$, followed by a peak around $w=7\sim 8$. Based on the overall performance, we select $w=6$ that achieves both efficiency and accuracy.

\section{Additional Statistics of User Study}
\label{appen:user_study}

\begin{table}[t]
\small
\centering
\caption{User study results}
\label{tab:statistics_user_study}
\resizebox{\columnwidth}{!}{
\begin{tabular}{c|ccc|ccc|cc|c}
\toprule
\textbf{P}
& \multicolumn{3}{c|}{\textbf{FHDR}}
& \multicolumn{3}{c|}{\textbf{UH-Random}}
& \multicolumn{2}{c|}{\textbf{Mech. Rate}}
& \textbf{Pref.} \\
\cmidrule(lr){2-4}
\cmidrule(lr){5-7}
\cmidrule(lr){8-9}
& A & S & T & D & A & T & F & S & \\
\midrule

1  & 30 & 20 & 147 & 3 & 4  & 57  & 6  & 4  & F \\
2  & 13 & 8  & 119 & 6 & 10 & 86  & 10 & 10 & U \\
3  & 0  & 0  & 0   & 2 & 4  & 59  & 4  & 7  & B \\
4  & 0  & 0  & 0   & 3 & 7  & 92  & 5  & 7  & B \\
5  & 2  & 1  & 15  & 3 & 4  & 72  & 6  & 6  & U \\
6  & 3  & 1  & 18  & 3 & 3  & 49  & 8  & 8  & F \\
7  & 16 & 9  & 74  & 4 & 8  & 124 & 9  & 7  & F \\
8  & 0  & 0  & 0   & 2 & 5  & 57  & 6  & 5  & U \\
9  & 30 & 12 & 198 & 4 & 7  & 98  & 7  & 7  & F \\
10 & 27 & 15 & 108 & 3 & 8  & 99  & 6  & 5  & B \\
11 & 13 & 6  & 98 & 4 & 8  & 102 & 7  & 6  & F \\
12 & 12 & 6  & 67  & 6 & 6  & 67  & 7  & 6  & F \\
13 & 0  & 0  & 0   & 3 & 6  & 91  & 7  & 3  & U \\
14 & 30 & 11 & 189 & 5 & 8  & 132 & 8  & 9  & U \\
15 & 9  & 6  & 90  & 3 & 6  & 64  & 8  & 8  & F \\
16 & 13 & 9  & 82  & 7 & 7  & 88  & 6  & 3  & F \\
17 & 9  & 6  & 52  & 2 & 5  & 77  & 7  & 6  & F \\
18 & 15 & 5  & 112 & 3 & 7  & 85  & 5  & 6  & B \\
19 & 10 & 5  & 45  & 4 & 6  & 99  & 5  & 8  & U \\
20 & 5  & 3  & 21  & 5 & 11 & 76  & 7  & 7  & F \\
21 & 33 & 21 & 182 & 3 & 8  & 117 & 9  & 4  & F \\
22 & 1  & 1  & 9   & 5 & 7  & 73  & 9  & 5  & F \\
23 & 31 & 21 & 110 & 3 & 9  & 112 & 10 & 5  & F \\
24 & 28 & 7  & 217 & 5 & 9  & 112 & 9  & 9  & F \\
25 & 32 & 15 & 174 & 2 & 7  & 77  & 4  & 8  & U \\
26 & 5  & 3  & 76  & 5 & 12 & 114 & 8  & 7  & F \\
27 & 28 & 9  & 176 & 3 & 9  & 119 & 7  & 5  & F \\
28 & 31 & 8  & 199 & 7 & 12 & 150 & 8  & 6  & F \\
29 & 0  & 0  & 0   & 4 & 7  & 86  & 8  & 8  & U \\
30 & 12 & 6  & 81  & 4 & 8  & 81  & 7  & 6  & F \\

\midrule
\textbf{AVG}
& 14.60
& 7.13
& 88.63
& 3.87
& 7.27
& 90.50
& 7.10
& 6.37
& $\substack{\mathrm{F}:18\\\mathrm{U}:8\\\mathrm{B}:4}$ \\

\bottomrule
\end{tabular}
}
\end{table}

We gathered $30$ participants for the user study, mostly university students. Detailed statistics from the user study are presented in Table \ref{tab:statistics_user_study}. The two main parts in the table, \textsc{FHDR} and \textsc{UH-Random}, records detailed statistics of a specific user using the algorithm. Due to limited space, we denote `P' as participant, `A' as the number of questions answered (including those being skipped), `S' as the number of skipped questions, `T' as the recorded time using the algorithm in seconds, including upfront attribute selection for \textsc{UH-Random}, and 'D' as the number of attributes selected for the \textsc{UH-Random} algorithm, as required. For FHDR sessions with no questions, $T=0$ denotes zero interaction time. Both algorithms contain only three columns: \textsc{FHDR} contains (A,S,T) as we do not know the number of attributes that the user considers implicitly while using the algorithm, and \textsc{UH-Random} contains (D,A,T) since skipping is not allowed. The remaining columns report the ratings of the two interaction mechanisms and the preference over recommendations returned by \textsc{FHDR} and \textsc{UH-Random}.

Upfront attribute-selection time for \textsc{UH-Random} was explicitly measured at approximately $50$~s and is included in its mean total time of $90.50$~s. FHDR’s mean recorded time is $88.63$~s. While the total times are similar, FHDR avoids upfront attribute selection and allows users to skip questions or stop answering at any time. Participants $3$, $4$, $8$, $13$, and $29$ requested recommendations without answering questions and received \textsc{AttributeSubset} outputs. Their $T=0$ entries denote zero interaction time. Other participants skipped questions during interaction, illustrating the use of FHDR’s flexible feedback options. Finally focusing on the preference over the final results, most participants favor the recommendation of our algorithm, which is either more accurately reflecting their preference, or could cover some preferred options successfully.